# Deformation transients of confined droplets within interacting electric and magnetic field environment[*]

Pulak Gupta,[†] Purbarun Dhar,[‡] and Devranjan Samanta[†]

A theoretical exploration and an analytical model for the electro-magneto-hydrodynamics (EMHD) of leaky dielectric liquid droplets, suspended in an immiscible confined fluid domain has been presented. The analytical solution for the system, under small deformation approximation, in creeping flow regime, has been put forward. Study of the droplet deformation suggests that its temporal evolution is exponential, and dependents on the electric and magnetic field interaction. Further, the direction of the applied magnetic field with respect to the electric field decides whether the contribution of magnetic forces opposes or aids the interfacial net electrical force due to the electric field. Validation of the proposed model at the asymptotic limits of vanishing magnetic field show that the model accurately reduces to the case of transient electrohydrodynamic model. We also propose a magnetic discriminating function ($\phi_M$) to quantify the steady-state droplet deformation in the presence of interacting electric and magnetic fields. The change of droplets from spherical shape to prolate, and oblate spheroids, correspond to $\phi_M > 0$ and $< 0$ regimes, respectively. It is shown that with the aid of low magnitude magnetic field, a substantial augmentation in the deformation parameter, and the associated EMHD circulation within and around the droplet is achieved. The analysis also reveals the deformation lag and specific critical parameters that aid or suppressed this lag behaviour; discussed in terms of relevant non-dimensional parameters.

## I. INTRODUCTION

The study of electric field-driven hydrodynamics at small scales, such as suspended liquid drops, is essential because of its relevance in wide range of industrial applications. A droplet, initially of spherical shape due to surface tension, may deform in the desired direction in presence of electric ($\vec{E}$) and magnetic field ($\vec{B}$), or their interactions. Many existing studies exclusively discuss the effect of electric field on the drop deformation dynamics, and associated electrohydrodynamics (EHD). The combined effect of electric and magnetic field (EMHD) could be more efficient for droplet manipulations than their corresponding individual effects. At the low Reynolds number regime, surface and capillary forces largely dominate the inertial forces. Consequently, it is exceedingly difficult to mix, manipulate, or transport droplets solely by hydrodynamic forces. In that sense, EMHD strategies that could be excellent solutions and can be exploited.

EHD study of droplets is useful in understanding of natural phenomena like the deformation and fragmentation behaviour of water droplets in thunderclouds [1]. Industrial applications such as EHD atomization is relevant for electro-spraying and electro-spinning [2]. Applications in the area of microfluidics, like manipulation of droplets in lab-on-a-chip by electro-wetting or electro-capillary [3], protein transfection into cells by electrical manipulation of droplets [4], the enhancement of heat or mass transfer to a drop by chaotic advection due to

EHD flows [5], inkjet printing [6, 7], and in EHD pumps, electrospray mass spectrometry, electrospray nanotechnology [7]. Other examples are phase separation by liquid-liquid extraction, dewatering of fine suspensions, and electrophoretic separation of biological materials [8]. There are certain applications where the efficiency of the process may be enhanced by imposing both electric and magnetic field, [9], EMHD micropump [10], and EMHD flow of blood [11], etc. In all these applications, droplet deformation is governed by the interplay of interfacial electrical, magnetic, and hydrodynamic stresses [5, 12–16].

Studies on "electro-hydrostatic theory" [17, 18] consider droplets as perfect dielectrics, or perfectly conducting liquids, suspended in an ideal dielectric surrounding fluid. In both cases, the droplet deforms as a prolate spheroid. It deforms in the electric field direction as theory predicts that net electrical stress is normal to the interface and directed from higher to lower values of permittivity. Allan and Mason [19] performed experiments and concluded that conducting droplets deform to prolate spheroid form; agreeing with the theory. They also showed that perfect dielectric droplets also deform perpendicular to the direction of the electric field to an oblate spheroid. Taylor [17] further showed that the oblate deformation is due to the interaction of hydrodynamic stresses along with electrical stresses at the interface. Due to this phenomenon, the bulk-free charges accumulate at the interface in presence of the electric field, and the possibility of net interfacial electrical stress arises [17, 20–22].

Droplet EHD was developed through Taylor [17]'s seminal work. He developed a mathematical framework for the steady-state axisymmetric EHD problem for an unconfined droplet in creeping flow regime, in the presence of low strength electric field ($E_o$). He showed that the ratio of electrical conductivity ($R = \sigma_{in}/\sigma_0$), where in the article the subscripts 'in' and 'o' represent the droplet and surrounding fluid, respectively) and the ratio of the electrical permittivity ($S = \epsilon_{in}/\epsilon_0$) plays a vital role in droplet deformation and EHD flow within the droplet and the surrounding. He observed that the electric field establishes rotational motion in the droplet, consisting of four equal strength vortices matching the corresponding vortices of the surrounding liquid. For R<S, the surrounding liquid flow from the poles to the equator, whereas flow reverses for R>S. For R=S, there will be no flow as the interface becomes charge free. The importance of the Capillary number (Ca $= \mu_o u_c/\gamma$, where $\mu_o$ is the dynamic viscosity of the fluid, $u_c$ is the velocity scale, and $\gamma$ is the coefficient of surface tension at the interface) was discussed. At low Ca, droplet deforms into ellipsoidal shape, and at higher Ca regime, droplet deviates from the ellipsoidal shape, and beyond a certain critical Ca, droplet disintegrates into smaller droplets [23].

---

[*] Corresponding author: devranjan.samanta@iitrpr.ac.in

[†] Department of Mechanical Engineering, Indian Institute of Technology Ropar, Punjab–140001, India

[‡] Hydrodynamics and Thermal Multiphysics Lab (HTML), Department of Mechanical Engineering, Indian Institute of Technology Kharagpur, West Bengal–721302, India

Next, Melcher and Taylor [20] reported the mathematical analysis for EHD under the influence of uniform finite electrical conductivity and permittivity. Leaky dielectric fluid was considered, where the system is not coupled with the electric field due to the absence of bulk charges. His pioneering study involves electromechanical coupling due to discontinuities of the electrical properties at the interfacial region, resulting in EHD motion. Vizika and Saville [15] found the expressions for steady-state deformation ($D_{ss}$) of the drop using Taylor's solution, by balancing normal interfacial stresses. Other works have used the deformation definition of Taylor [24] and predicted the end shape of the droplet from the sign of the Taylor discriminating function ($\phi$) [16], and its magnitude is proportional to the applied electric field strength ($E_o^2$). Other studies based on similar grounds report leaky dielectric droplet dynamics[17, 19, 20, 22, 25–30] considering steady state EHD and without confinements.

Transient EHD of droplets [12, 16, 31–33] and the confinement of the droplet [13] have been explored sparsely in literature. Esmaeeli and Sharifi [31] developed a mathematical model for the temporal evolution of leaky dielectric spherical droplets and distortion from a circular shape for a leaky dielectric liquid column [12] during EHD. Later, Esmaeeli and Behjatian [13] also considered confined domains in their study, a condition where droplet and surrounding fluid dimensions are comparable. Their analysis show that in a confined domain R>1 system, the electric field strengthens and flow field intensity either remains the same or increases; and similarly, for R<1, the electric field and flow field intensity decrease. Later, Mandal et al. [34] considered both transient and confinement effects with finite charge relaxation time in their analysis. They found that the interfacial charging process consumes less time in a confined domain than in an unconfined domain. In transient EHD, liquid drops deform monotonically when the droplet interface charges rapidly, and non-monotonically when charge relaxation is finite [35, 36].

Tsukada et al. [37] studied the effect of inertia (using Galerkin finite element calculations) and compared their numerical and experimental results with Taylor's results. They show that as the electric field strength increases, the asymptotic solution of Taylor is no longer valid as inertia also acts. Feng [38] further showed that induced flow intensity reduces due to interfacial free charge convection. Additionally, they explored that oblate droplets deform less when the charge convection is present. The opposite is true for prolate shape droplets, where charge convection aids deformation. Works show that as the electric field strength increases, droplets may become unstable and disintegrate above a critical field strength [18, 39]. Sherwood [40] performed experiments and numerical simulations to study the effect of electric fields on the droplet breakup process in the surrounding fluid. Two modes of breakup were noted: (a) when the S>1, then small daughter droplets stream away from the droplet forming pointed filaments, and (b) division of the droplet into two large daughters, connected by thin liquid thread when the R>1. Lac and Homsy [28] extended Sherwood [40]'s work through simulations to identify different breakup modes and the effect of the viscosity of the fluids. Later Bentenitis and Krause [41] proposed an extended leaky dielectric method (EDLM) to analyse highly deformed droplets under strong electric field intensity. They suggest two modes of deformation at a high electric field: continuous deformation or hysteresis.

A comprehensive study of the literature reveals that an understanding of droplet dynamics under the simultaneous effect of the electric and magnetic fields (EMHD) is unavailable. Further, though attention has been given to the steady-state and transient analysis of a droplet EHD in confined or unconfined domain, most studies ignore the complexities associated with the Lorentz force. The impact of EMHD would be especially important for droplets or surroundings with high electrical conductivities, such as strong electrolytes, ionic liquids, and liquid metals. Orthogonal electric and magnetic fields would result in a net driving body force to stimulate the internal and surrounding hydrodynamics and the droplet deformation kinetics. Based on the above motivation, explore the problem of transient EMHD of a droplet in a generalized confined domain, and portray its mathematical nature. We determine the temporal evolution of droplet deformation and EMHD flow field under the small deformation approximation. We validate our droplet deformation and flow results at the asymptotic limits of zero magnetic field; and are able to recover the EHD behaviour as shown in literature. We have also performed a detailed analysis of droplet deformation parameters by varying electric field strength keeping magnetic field strength constant, and vice versa. Finally, we show the influence of strength and direction of applied magnetic field on the variation of droplet deformation parameter and on the streamline patterns of the EMHD flow compared to pure EHD scenario.

## II. DESCRIPTION OF THE PHYSICAL SYSTEM

Figure 1 shows the schematic of the physical problem. The spherical dielectric liquid droplet of radius $R_{in}$ is suspended in another immiscible dielectric liquid medium, confined by a rigid spherical container of radius $R_o$. The immersed droplet and the container are concentric at the origin. We have adopted the spherical coordinate system $(r, \theta, \phi)$ to analyze the problem. The fluids are neutrally buoyant and stationary due to symmetry, and the center of mass of the system remains at the origin. As we are assuming an axisymmetric spherical coordinate system, the flow field within the droplet and in the confined domain remain axisymmetric to the direction of applied uniform electric field $E_o$ (along y-axis) and steady uniform magnetic field (B) perpendicular to the $E_o$. Hence, our problem is independent of the azimuthal direction $(\phi)$ [14, 17, 31]. However, in case of strong magnetic field, this assumption may not be valid and induced magnetic field may not be negligible [42].

Our study is restricted to a small intensity of the electric field, so the axisymmetric assumption is valid. The deformation due to the electrical, magnetic, and hydrodynamic stresses is also influenced by the confining domain. Based on previous studies [43–46], we have implemented certain assumptions for further simplification of the mathematical model, as:

1. Fluids are assumed to be Newtonian. The physical properties of the fluids are time invariant and also due to application of electric and magnetic field.

2. Initially, the droplet shape is spherical. Both droplet and confined surrounding liquid are concentric throughout.

3. Applied electric field is of low strength such that there is no induced additional magnetic field. The electric field is irrotational (i.e., $\vec{\nabla} \times \vec{E} = \vec{0}$), and the flow field is axisymmetric to it.

4. The electric Reynolds number ($Re_E = \varepsilon_o u_c / \sigma_o R_{in} \ll 1$) signifies the relative order of charge convection to the Ohmic conduction. This leads to negligible charge relaxation time compared to charge convection.

5. In case of the leaky dielectrics, on application of electric field, charges from bulk liquid move to the droplet interface. Hence the charge density in the bulk fluid is zero, but non-zero at the interface.

6. The external applied magnetic field across the drop interface is continuous and does not change as the drop deforms.

The physical properties of the illustrative droplet and confined liquids have been tabulated in II. Relevant ratios used are R, S and $\lambda = \mu_{in}/\mu_0$ (the dynamic viscosity ratio)

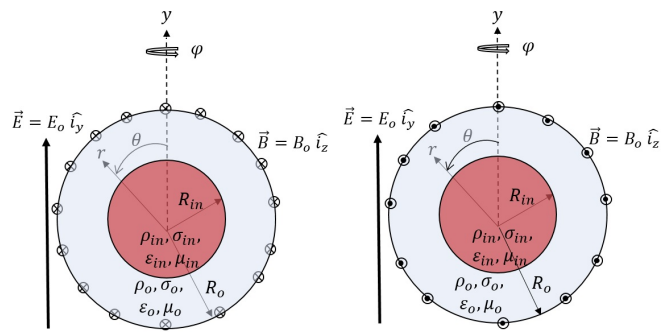

FIG. 1. Schematic of a suspended droplet (radius $R_{in}$) in a spherical confined geometry (radius $R_o$), depicted in a spherical coordinate system $(r, \theta, \phi)$. The direction of the applied uniform direct electric field is the y-direction. The direction of the applied uniform steady magnetic field is perpendicular to the $E_o$ and directed (a) inward and (b) outward from the plane of paper.

[17]. For non-dimensionalization, the radius of the drop is selected as the length scale ($l_c \sim R_{in}$), the velocity scale ($u_c \sim \epsilon_o E_o^2 R_o / \mu_o$) is obtained by balancing electrical and viscous stresses, and the time scale ($t_c \sim \mu_o R_o / \gamma$) denotes the time the droplet takes to reach an equilibrium shape after field is applied. This includes charge relaxation time and shape relaxation time scales. The charge relaxation time is the time taken by all the charges to reach the interface, and the shape relaxation time is the time taken by the droplet to reach a stable, deformed shape. The other relevant parameters are the pressure scale ($p_c \sim \mu_o u_c / R_o$), electric potential scale ($\phi_c \sim E_o R_o$), and surface charge density scale ($q_c \sim \varepsilon_o E_o$).

## III. GOVERNING EQUATIONS WITH BOUNDARY CONDITIONS

### A. Governing equation for the electric potential and electric field distribution

In this analysis, conservation of mass, momentum, charge, and Maxwell's electromagnetic equations are relevant. Since the surrounding liquid has constant properties and no charge in the bulk fluid (leaky dielectric model), the electric field equations may be decoupled from the flow equations. Thereby, the hydrodynamics equations are solved via momentum jump condition [20, 21, 48] across the interface. The approach is used to solve a transition jump in normal and tangential stresses across the interface induced by the electric and magnetic field and property mismatch. The electric field intensity is irrotational and divergence-free (for leaky dielectric with a small dynamic electric current). The irrotational nature ($\vec{\nabla} \times \vec{E} = \vec{0}$), together with Faraday's law, ($\vec{\nabla} \times \vec{E} = -\frac{\partial \vec{B}}{\partial t}$) implies that that the induced magnetic field is negligible. Further, the electric field can be ex-pressed as a gradient, $\overrightarrow{E} = -\overrightarrow{\nabla}\phi$. The small dynamic electric current [20, 21] condition leads to the Laplace condition for the electric potential [20] as:

$$\overrightarrow{\nabla}^2\phi = 0 \qquad (1)$$

TABLE I. Physical properties of the fluids used. Here, $R_{in} = 10$ mm, $E_o = 10$ V/mm, $\gamma = 5.5 \times 10^{-3}$ [$Nm^{-1}$] and $\epsilon_o = 8.854 \times 10^{-12}$ [F-$m^{-1}$]. System A correspond to oxidized castor oil droplet surrounded by silicon oil. System B is mentioned in Table 1. of supplementary section and is obtained by interchanging the droplet and surrounding fluid of system A [31, 47].

| System A | $\sigma$(S/m) | $\varepsilon/\varepsilon_0$ | $\mu$(Kg/m−s) | $\rho$ (Kg/m³) |
|---|---|---|---|---|
| Droplet (oxidized castor oil) | $1.11 \times 10^{-12}$ | 6.3 | 6.5 | 980 |
| Suspending Medium (silicon oil) | $3.33 \times 10^{-11}$ | 2.77 | 12 | 980 |

For a more generalized study, we have considered two different systems:

1. A perfect leaky system, where both droplet and surrounding liquid are leaky dielectrics with small finite electrical conductivity. The electric potential distribution is obtained by solving $\overrightarrow{\nabla} \cdot (\sigma\overrightarrow{\nabla}\phi) = 0$ [43, 49, 50]. This results from the conservation of electric charge, $(\nabla \cdot J + \frac{Dq_v}{Dt} = 0)$, where J is free current density.

2. A perfect dielectric system, where the two fluids are ideal dielectric. The Gauss law is applicable, which relates the electric displacement of free charges to the volumetric charge density $(q_v)$ of a system having permittivity $(\epsilon)$ as $\overrightarrow{\nabla} \cdot (\varepsilon\overrightarrow{\nabla}\phi) = q_v$ [51]. For a perfect dielectric system, the bulk charge density is zero. Hence the electric potential distribution is obtained from $\overrightarrow{\nabla} \cdot (\varepsilon\overrightarrow{\nabla}\phi) = 0$.

Equation 1 is constrained in the present case by the following boundary and interfacial conditions:

(a1) $\phi_{in}(0,\theta)$ is bounded, i.e., electric potential within the droplet should be finite.

(a2) The electric potential should be continuous across the interface, i.e. $\phi_{in}(1,\theta) = \phi_o(1,\theta)$. This leads to in symmetric distribution of electric field in azimuthal direction throughout.

(a3) $[\![-\varepsilon\overrightarrow{\nabla}\phi.\hat{n}]\!] = q(t,\theta)$, i.e., charge conduction and convection balance one another. This results in the non-symmetric distribution of the electric field in a normal direction within the system and leads to distribution of charge density across the interface. Here $\hat{n}$ is the normal outward vector to the droplet surface, $[\![Q]\!] = Q_o - Q_{in}$ denotes the jump in any generic parameter Q across the interface.

(a4) Electric potential near the wall of the spherical container varies as $\phi_o(1/\alpha,\theta) = \frac{-\cos\theta}{\alpha}$, where $\alpha$ is the confinement ratio (defined as the ratio of $R_{in}$ and $R_o$).

Due to electric and magnetic fields, the interface of the droplet will not be perfectly spherical and will deform. To quantify the temporal evolution of deformation, a deformation parameter is proposed as: $|D| = \frac{L-B}{L-B}$, where

L and B are the lengths of axis oriented parallel to the direction of the applied electric field, and perpendicular to the direction of the applied electric field, respectively. Our small deformation scenario suggests $|D| \ll 1$.

### B. Governing equation for charge conservation at the interface

The distribution of surface charge density is estimated from charge conservation across the droplet interface, which is caused due to a jump in electric displacement across the interface. Electric charge conservation is a fundamental principle and will not be affected by the applied external magnetic field. The magnetic field can alter the motion of the system's charged particle but can't destroy or create charges. The charges move from the bulk fluid to the interface due to electric potential gradients or carried by the flow. The charge transportation for leaky dielectric liquids is small in the external magnetic field. These liquids have low electrical conductivity, permittivity, permeability and induced current due to the interaction of externally applied electric and magnetic fields. Further, the mathematical analysis compares the charge relaxation time of the electrical and magnetic phenomena.

Relaxation time for the electrical phenomena $(\tau_E) = \frac{\text{electrical permittivity } (\varepsilon)}{\text{electrical conductivity } (\sigma)}$

Relaxation time for the electrical phenomena $(\tau_B) =$ magnetic permeability $(\mu) \times$ conductivity $(\sigma) \times l_c^2$

By using the properties of the system, $\tau_E = O\left(10^{-3}\text{sec}\right)$ & $\tau_B = O\left(10^{-12}\text{sec}\right)$, i.e. $\tau_E \gg \tau_B$. It can be observed that charge transportation due to magnetic phenomena is much less affected than due to electrical phenomena. Once all the charges reach the interface, then the magnetic effect coupled with the electrical effect affects the deformation and the flow field analysis.

The non-dimensional form of charge conservation across the interface (ignoring lateral surface diffusion) [21, 33] is as:

$$\frac{Sa}{Oh^2}\frac{\partial q}{\partial t} + Re_E\left[\vec{u} \cdot \overrightarrow{\nabla_s}q - q\hat{n} \cdot (\hat{n} \cdot \overrightarrow{\nabla})\vec{u}\right] = [\![-\sigma\overrightarrow{\nabla}\phi]\!] \cdot \hat{n} \quad (2)$$

where, $\overrightarrow{\nabla_s} = \overrightarrow{\nabla} - \hat{n} \cdot (\hat{n} \cdot \overrightarrow{\nabla})$ is the surface divergence, and non-dimensional numbers are Sa $= \varepsilon_o u_o/\sigma_o R_{in}^2$ (a form of ratio of the electric Stokes number to Electroviscous number), Ohnesorge Number (Oh $= \mu_o/\sqrt{\rho_o R_{in}\gamma}$, relates the viscous force to inertia and surface tension force) and $\mathrm{Re_E}$. Mathematically these non-dimensional numbers can be expressed as the ratio of various time scales as: $\frac{\tau_E}{\tau_p} = \frac{Sa}{Oh^2} = \frac{\varepsilon_o/\sigma_o}{\mu_o R_{in}/\gamma}$ and $\frac{\tau_E}{\tau_f} = \mathrm{Re_E} = \frac{\varepsilon_o/\sigma_o}{\mu_o/u_c}$. Here, $\tau_E = \varepsilon_o/\sigma_o =$ characteristic time for charge relaxation. A small value of $\tau_c$ suggests rapid charging of the interface Ha and Yang [52].

$\tau_p = \mu_o R_{in}/\gamma =$ surface deformation time scale or capillary time scale

$\tau_f = R_{in}/u_c =$ convection flow time scale

$\tau_\mu = R_{in}^2/v_o =$ time scale of momentum diffusion by viscosity

In LHS of Equation 2, $1^{st}$ term depicts displacement current. This term plays a role in very low electrical conductivities when interface charging time becomes comparable with the deformation time scale. In contrast, $2^{nd}$ term shows convection of surface charge at the interface due to fluid flow and dilation, while the RHS term shows Ohmic conduction normal to the interface. Applying order of magnitude analysis using properties from Table II, we found that the $2^{nd}$ term of the LHS is negligible compared to the others. Consequently the Equation 2 is reduced as:

$$\frac{Sa}{Oh^2}\frac{\partial q}{\partial t} = [\![ -\sigma \vec{\nabla}\phi ]\!] \cdot \hat{n} \tag{3}$$

### C. Governing equation for EMHD

The non-dimensional continuity equation for incompressible flow is used:

$$\vec{\nabla} \cdot \vec{u} = 0 \tag{4}$$

The Navier-Stokes equation with effects of applied electric $(\overrightarrow{f_E})$ and magnetic $(\overrightarrow{f_M})$ body force density is as:

$$\rho\left(\frac{\partial \vec{u}}{\partial t} + \vec{u}\cdot\vec{\nabla}\vec{u}\right) = -\vec{\nabla}p + \mu\vec{\nabla}^2\vec{u} + \overrightarrow{f_E} + \overrightarrow{f_M} \tag{5}$$

The divergence of the Maxwell stress tensor gives the expression for the electrical body force density exerted, as:

$$\overrightarrow{f_E} = q_v E - \frac{1}{2}E^2\nabla\varepsilon + \frac{1}{2}\nabla\left[\rho\left(\frac{\partial\varepsilon}{\partial\rho}\right)_T E^2\right] \tag{6}$$

Where, $1^{st}$ term of RHS represents Coulomb force on the free charges in the bulk fluid due to the imposed electric field. The $2^{nd}$ term represents polarization due to non-uniformity of the electrical permittivity. The $3^{rd}$ term signifies electrostrictive force arising due to variation of electrical permittivity with density at a fixed temperature. As the permittivity is constant the $3^{rd}$ term can be neglected. The $2^{nd}$ term is non-zero only at the interface of the two fluids. For the leaky dielectric model, free charge density in the bulk fluid is zero. But free charge density is not zero at the interface. So due to impose electric field, the electrical forces are non-zero at the interface. These electrical forces at the interface cause electrohydrodynamic coupling.

The magnetic force $(\overrightarrow{f_M})$ can be evaluated from Lorentz force under assumption of an irrotational applied electric field, as:

$$\overrightarrow{f_M} = \vec{J} \times \vec{B} \tag{7}$$

Here, J is free electric current density, related to the electric field by Ohm's law as $\vec{J} = \sigma(\vec{E} + \vec{u}\times\vec{B})$. Incorporating the expressions of $(\overrightarrow{f_E})$ and $(\overrightarrow{f_M})$, Equation 5 yields:

$$\frac{Re}{Ca}\left(\frac{\partial\vec{u}}{\partial t}\right) + \mathrm{Re}\,\vec{u}\cdot\vec{\nabla}\vec{u} = -\vec{\nabla}p + \lambda\vec{\nabla}^2\vec{u} + E_v\cdot Ha\vec{E} - Ha^2\vec{u}$$

Where, $E_v(\text{ Electroviscous number }) = \frac{E_o R_{in}}{u_c}\sqrt{\frac{\sigma_o}{\mu_o}}$,

$$\mathrm{Ha(Hartmann\ Number)} = \sqrt{\frac{\sigma_0 B^2}{\mu_0/R_{in}^2}}$$

The above equation can be expressed in terms of Oh as:

$$Oh^{-2}\left(\frac{\partial\vec{u}}{\partial t}\right) + Re\vec{u}\cdot\vec{\nabla}\vec{u} = -\vec{\nabla}p + \lambda\vec{\nabla}^2\vec{u} + E_v\cdot Ha\vec{E} - Ha^2\vec{u} \tag{8}$$

Since, $Oh^{-2} << 1$ implies that the momentum diffusion time scale $(\tau_\mu)$ is much smaller than droplet surface deformation time scale $(\tau_c)$. By this argument, $1^{st}$ term of LHS can be neglected. Further, by order of magnitude analysis, the $2^{nd}$ term of LHS is negligible in creeping flows, and the $3^{rd}$ term of RHS can also be neglected as the electromagnetic effect is insignificant compared to the viscous effect due to non-availability of free charge density in the bulk fluid. Since induced current due to interaction of electric and magnetic field is negligible, the $4^{th}$ term of RHS of the equation can be neglected. All such effects are significant at the interface, and shall be brought in later separately into the formulation. Therefore, Equation 8 is further simplified to

$$-\vec{\nabla}p + \lambda\vec{\nabla}^2\vec{u} = \vec{0} \tag{9}$$

By taking the curl, the Equation 9 is transformed into:

$$\nabla^2\vec{\omega} = 0 \tag{10}$$

where,

$$\omega = \frac{1}{r}\left[\frac{\partial}{\partial r}(ru_\theta)\right] - \frac{1}{r}\left[\frac{\partial}{\partial\theta}(u_r)\right] \tag{11}$$

Where, the radial velocity $u_r = \frac{1}{r^2\sin\theta}\frac{\partial\psi}{\partial\theta}$ and tangential velocity $u_\theta = \frac{-1}{r\sin\theta}\frac{\partial\psi}{\partial r}$ and $\psi$ represents the stream function. By substituting the value of $u_r$ and $u_\theta$ in Equation 11, we get

$$\nabla^2\psi = -\omega \tag{12}$$

By solving Equation 10 and Equation 12, we get

$$\nabla^2\left(\nabla^2\psi\right) = \nabla^4\psi = 0 \tag{13}$$

Where operator $\nabla^2$ represents

$$\nabla^2 \equiv \frac{\partial^2}{\partial r^2} + \left(\frac{\sin\theta}{r^2}\right)\left[\frac{\partial}{\partial\theta}\left\{\left(\frac{1}{\sin\theta}\right)\frac{\partial}{\partial\theta}\right\}\right].$$

The boundary and interfacial conditions for the above equation are as:

(b1) Velocity should be bounded at the center of the droplet, i.e., $u_{in,r}(0,\theta)$ and $u_{in,\theta}(0,\theta)$ should be bounded.

(b2) No-slip boundary condition at the droplet interface i.e. $u_{in,\theta}(1,\theta) = u_{o,\theta}(1,\theta)$.

(b3) No normal flow at the droplet interface i.e. $u_{in,r}(1,\theta) = u_{o,r}(1,\theta) = \frac{1}{Ca}\left[\frac{d\xi}{dt}\right]_{r=1}$ where, $\xi$ is a non-dimensional droplet shape function that connects the continuity of radial components of velocities for the deformed interface.

(b4) Velocity at the outer confining wall should be zero, i.e., $u_{o,r}(1/\alpha,\theta) = u_{o,\theta}(1/\alpha,\theta) = 0$

(b5) $[\![\tau_{r\theta}^T]\!] = [\![\tau_{r\theta}^H]\!] + [\![\tau_{r\theta}^E]\!] + [\![\tau_{r\theta}^B]\!] = 0$; signifying the balance of a jump in tangential shear stresses due to EMHD, to ascertain no shape mismatch at the interface.

(b6) $[\![\tau_{rr}^T]\!] = [\![\tau_{rr}^H]\!] + [\![\tau_{rr}^E]\!] + [\![\tau_{rr}^B]\!] - [\![p]\!] = \frac{\kappa}{Ca}$; signifying the balance of a jump in normal shear stresses due to EMHD with the interfacial tension, to ensure no normal flow across the interface. $\kappa$ is non-dimensional interfacial curvature by imposing normal stress balance. Here superscripts H, E, B and T denote hydrodynamic, electric, magnetic and total components. For $|D| \ll 1$, the drop shape function is adopted from reports [15, 31] as:

$$\xi = \left[1 + \frac{2D}{3}\left(3\cos^2\theta - 1\right)\right] \tag{14}$$

And the local interfacial curvature as:

$$\kappa = \left[\frac{8D}{3}\left(3\cos^2\theta - 1\right) + 2\right] \tag{15}$$

## IV. SOLUTION OF THE MATHEMATICAL DESCRIPTION

### A. Solution for hydrodynamics and jump in hydrodynamic stresses

Hydrodynamic equations were solved by using the stream function as $\psi = r^n \sin^2\theta\cos\theta$, as suggested by boundary condition (b5), where n is a real-valued constant. Substituting $\psi$ in Equation 13 leads to a solution n = 0, –2, 3, and 5. Finally, we can write the stream function for droplet ($\psi_1$) and surrounding liquid domain ($\psi_2$) as:

$$\psi_1 = \left[A_{in} + B_{in}r^{-2} + C_{in}r^3 + D_{in}r^5\right]\sin^2\theta\cos\theta \tag{16}$$
$$\psi_2 = \left[A_o + B_or^{-2} + C_or^3 + D_or^5\right]\sin^2\theta\cos\theta \tag{17}$$

Where $A_{in}$, $B_{in}$, $C_{in}$, $D_{in}$, $A_o$, $B_o$, $C_o$, $D_o$ are constants will obtain from conditions (b1) – (b6). From (b1), $A_{in} = B_{in} = 0$, and thus we get

$$\psi_1 = \left[C_{in}r^3 + D_{in}r^5\right]\sin^2\theta\cos\theta \tag{18}$$

Condition (b2) yields

$$3C_{in} + 5D_{in} = -2B_o + 3C_o + 5D_o \tag{19}$$

Next, (b3) yields

$$Ca\left(C_{in} + D_{in}\right) = \frac{2}{3}\frac{dD}{dt} \tag{20}$$

and,

$$Ca\left(A_o + B_o + C_o + D_o\right) = \frac{2}{3}\frac{dD}{dt} \tag{21}$$

And (b4) yields

$$A_o\alpha^5 + B_o\alpha^7 + C_o\alpha^2 + D_o = 0 \tag{22}$$

and,

$$-2B_o\alpha^7 + 3C_o\alpha^2 + 5D_o = 0 \tag{23}$$

From Equation 19 to Equation 23, we write $A_o$, $C_o$, $D_o$, $C_{in}$, $D_{in}$ in terms of $B_o$ as:

$$
\begin{aligned}
A_o &= \frac{1}{f_1}\left[\frac{4}{3Ca}\frac{dD}{dt} - B_of_2\right] \\
C_o &= \frac{\alpha^3}{f_1}\left[-\frac{10}{3Ca}\frac{dD}{dt} + B_of_3\right] \\
D_o &= \frac{\alpha^5}{f_1}\left[\frac{2}{Ca}\frac{dD}{dt} - B_of_4\right] \\
C_{in} &= \frac{1}{2f_1}\left[\frac{20}{3Ca}\frac{dD}{dt}\left(1 - \alpha^3\right) - B_of_5\right] \\
D_{in} &= \frac{1}{2f_1}\left[\frac{4}{Ca}\frac{dD}{dt}\left(\alpha^5 - 1\right) + B_of_5\right]
\end{aligned}
\tag{24}
$$

Where $f_1$, $f_2$, $f_3$, $f_4$, and $f_5$ are dependent functions of $\alpha$ and are discussed in Appendix A.

To impose conditions (b4) and (b5), the velocity field, distribution of electric potential, surface free charge density, electrical stresses, and magnetic stresses in the domain should be known. A jump in the shear and normal hydrodynamic stress, and jump in pressure across the interface can be obtained from these velocities.

Jump in hydrodynamic normal stress:

$$
\begin{aligned}
[\![\tau_{rr}^H]\!] =& 2\frac{\partial u_{2r}}{\partial r} - 2\lambda\frac{\partial u_{1r}}{\partial r} \\
=& \left[(4A_o + 8B_o - 2C_o - 6D_o) + \lambda(2C_{in} + 6D_{in})\right]\left(1 - 3\cos^2\theta\right)
\end{aligned}
\tag{25}
$$

Jump in hydrodynamic shear stress:

$$
\begin{aligned}
[\![\tau_{r\theta}^H]\!] =& \left[r\frac{\partial}{\partial r}\left(\frac{u_{2\theta}}{r}\right) + \frac{1}{r}\frac{\partial u_{2r}}{\partial\theta}\right] - \lambda\left[r\frac{\partial}{\partial r}\left(\frac{u_{1\theta}}{r}\right) + \frac{1}{r}\frac{\partial u_{1r}}{\partial\theta}\right] \\
=& \left[(-6A_o - 16B_o - 6C_o - 16D_o) + \lambda(6C_{in} + 16D_{in})\right]\sin\theta\cos\theta
\end{aligned}
\tag{26}
$$

Jump in pressure:

$$
\begin{aligned}
[\![p]\!] =& p_2 - \lambda p_1 = \int\left(\frac{\partial p_2}{\partial r} - \lambda\frac{\partial p_1}{\partial r}\right)dr \\
=& \left[(2A_o + 7D_o) - 7\lambda D_{in}\right]\left(3\cos^2\theta - 1\right) + \text{ constant}
\end{aligned}
\tag{27}
$$

Where, $\frac{\partial p}{\partial r} = \mu\left[\frac{\partial}{\partial r}\left(\frac{1}{r^2}\frac{\partial}{\partial r}\left(r^2 u_r\right)\right) + \frac{1}{r^2\sin\theta}\frac{\partial}{\partial\theta}\left(\sin\theta\frac{\partial u_r}{\partial\theta}\right) - \frac{2}{r^2\sin\theta}\frac{\partial}{\partial\theta}\left(u_\theta\sin\theta\right)\right]$

## B. Solution for distribution of electric field and jump in electrical stresses

To obtain the electric field distribution, we begin with the solution of Equation 1 in axisymmetric form. Separation of variables is performed on Equation 1 as:

$$
\phi(r,\theta) = R(r)\Theta(\theta)
\tag{28}
$$

Here R(r) is the function of r only, and $\Theta(\theta)$ is the function of $\theta$ only. We get:

$$
\frac{\partial\phi}{\partial r} = R'(r)\Theta(\theta) \quad \& \quad \frac{\partial\phi}{\partial\theta} = R(r)\Theta'(\theta)
\tag{29}
$$

Substituting the values from Equation 29 into Equation 1, we get:

$$
\vec{\nabla}^2\phi = \frac{1}{R(r)}\frac{d}{dr}\left(r^2 R'(r)\right) + \frac{1}{\theta(\theta)\sin\theta}\frac{d}{d\theta}\left(\sin\theta\Theta'(\theta)\right) = 0
\tag{30}
$$

From separating the variable, Equation 30 yields:

$$
\frac{1}{R(r)}\frac{d}{dr}\left(r^2 R'(r)\right) = m(m+1)
\tag{31}
$$

$$
\frac{1}{\Theta(\theta)\sin\theta}\frac{d}{d\theta}\left(\sin\theta\Theta'(\theta)\right) = -m(m+1)
\tag{32}
$$

Equation 32 can be written as:

$$
\frac{d^2\Theta}{d\theta^2} + \frac{\cos\theta}{\sin\theta}\frac{d\Theta}{d\theta} + m(m+1)\Theta = 0
\tag{33}
$$

Equation 32 can be solved by the method of Frobenius to yield the solution:

$$
R(r) = Ar^m + Br^{-(m+1)}
\tag{34}
$$

Using $x = cos\theta$ and applying the chain rule in Equation 33 converts it to the form of Legendre equation $\left[\left(1 - x^2\right)\frac{d^2 y}{dx^2} - 2x\frac{dy}{dx} + m(m+1)y = 0\right]$. The general solution to the Legendre equation is the Legendre polynomials, $P_m(\cos\theta)$, and the complete general solution of the Laplace equation in axisymmetric spherical coordinates is as:

$$
\phi(r,\theta) = \sum_{m=0}^{\infty}\left(A_m r^m + B_m r^{-(m+1)}\right)P_m(\cos\theta)
\tag{35}
$$

Where, $P_m(\cos\theta)$ is $m^{\text{th}}$ order polynomial. By using condition (a1), (a2) and (a4) in Equation 35 we obtain the electric potential function as [6]:

$$
\begin{aligned}
\phi_1(r,\theta) =& \left[-r + d_1 r\left(1 - \alpha^3\right)\right]P_1 + \sum_{m=2}^{\infty} d_m r^m\left(1 - \alpha^{2m+1}\right)P_m(\cos\theta) \\
\phi_2(r,\theta) =& \left[-r\left(1 + d_1\alpha^3\right) + d_1 r^{-2}\right]P_1 + \sum_{m=0}^{\infty}\left(-d_m r^m\alpha^{2m+1} + d_m r^{-m-1}\right)P_m(\cos\theta)
\end{aligned}
\tag{36}
$$

Now, using condition (a3) in $\phi_1$ and $\phi_2$ of Equation 36, we can rewrite (a3) as:

$$q(t,\theta) = [\![-\varepsilon\vec{\nabla}\phi.\hat{n}]\!] = \left[S\vec{\nabla}\phi_{\text{in}} - \vec{\nabla}\phi_o\right].\hat{n}$$

And hence,

$$
\begin{aligned}
q(t,\theta) &= \left[S\left\{-1+d_1\left(1-\alpha^3\right)\right\} + \left(1+d_1\alpha^3\right) + 2d_1\right]P_1, \text{ and} \\
q(t,\theta) &= \left[Smr^{m-1}d_m\left(1-\alpha^{2m+1}\right) - d_m\alpha^{2m+1}mr^{m-1} + d_m(-m-1)r^{-m-2}P_m, \forall m \geq 2\right.
\end{aligned}
\tag{37}
$$

Now, invoking Equation 36 and Equation 37 with equation Equation 3, we get

$$
\begin{aligned}
\frac{Sa}{Oh^2}\left[\frac{(S+2)-\alpha^3(S-1)}{(R+2)-\alpha^3(R-1)}\right]\frac{d}{dt}\left(d_1\right) &= \frac{(R-1)}{(R+2)-\alpha^3(R-1)} \\
\frac{Sa}{Oh^2}\left[mS\left(1-\alpha^{2m+1}\right) + m\alpha^{2m+1} + m+1\right]\frac{d}{dt}\left(d_m\right) &= \\
&-\left[mR\left(1-\alpha^{2m+1}\right) + m\alpha^{2m+1} + m+1\right]d_m
\end{aligned}
\tag{38}
$$

Here, $\tau_1 = \frac{Sa}{Oh^2}\left[\frac{(S+2)-\alpha^3(S-1)}{(R+2)-\alpha^3(R-1)}\right]$ is form of non-dimensional charge relaxation time scale [6].

Solving Equation 38 with the initial condition of charge free droplet interface i.e., q $(0,\theta)$ is zero gives

$$
\begin{aligned}
d_1 &= \left(\frac{R-1}{R+2}\right)\left[\frac{(R+2)}{(R+2)-\alpha^3(R-1)}\right] \\
&+ \left(\frac{3(S-R)}{(R+2)(S+2)}\right)\left[\frac{(R+2)}{(R+2)-\alpha^3(R-1)}\right]\left[\frac{(S+2)}{(S+2)-\alpha^3(S-1)}\right]e^{-\frac{t}{\tau_1}} \\
d_m &= 0, \forall m \geq 2
\end{aligned}
\tag{39}
$$

Now, the closed-form distribution function of electric potential is obtained as:

$$
\begin{aligned}
\phi_{\text{in}}\,(r,\theta) &= \frac{-3\Gamma_R r}{R+2}\left[1 - \frac{\Gamma_S(S-R)\left(1-\alpha^3\right)}{S+2}e^{-\frac{t}{\tau_1}}\right]\cos\theta \\
\phi_o(r,\theta) &= -\Gamma_R\cos\theta\left[\left(r - \frac{R-1}{R+2}\frac{1}{r^2}\right) + \frac{3\Gamma_S(S-R)}{(R+2)(S+2)}\left(r\alpha^3 - \frac{1}{r^2}\right)e^{-\frac{t}{\tau_1}}\right]
\end{aligned}
\tag{40}
$$

where, $\Gamma_R = \frac{(R+2)}{(R+2)-\alpha^3(R-1)}$ and $\Gamma_S = \frac{(S+2)}{(S+2)-\alpha^3(S-1)}$ are termed as confinement correction factors. The expression for distribution of the electric field can be obtained considering irrotational and divergence-free field, i.e.

$$\vec{E} = -\vec{\nabla}\phi = -\left[\frac{\partial\phi}{\partial r} + \frac{1}{r}\frac{\partial\phi}{\partial\theta}\right]\tag{41}$$

Now, from Equation 40 and Equation 41, the electric field distribution within the droplet and in the surrounding domain is obtained as:

$$\vec{E}_{\text{in}} = \frac{3\Gamma_R}{(R+2)}\left[1 - \frac{\Gamma_S(S-R)\left(1-\alpha^3\right)}{(S+2)}e^{-\frac{t}{\tau_1}}\right]\left(\cos\theta\widehat{i_r} - \sin\theta\widehat{i_\theta}\right)\tag{42}$$

$$
\begin{aligned}
\vec{E}_o &= \Gamma_R\cos\theta\left[1 + \frac{2(R-1)}{(R+2)}\frac{1}{r^3} + \frac{3\Gamma_S(S-R)}{(R+2)(S+2)}\left(\alpha^3 + \frac{2}{r^3}\right)e^{-\frac{t}{\tau_1}}\right]\widehat{i_r} \\
&- \Gamma_R\sin\theta\left[1 - \frac{(R-1)}{(R+2)}\frac{1}{r^3} + \frac{3\Gamma_S(S-R)}{(R+2)(S+2)}\left(\alpha^3 - \frac{1}{r^3}\right)e^{-\frac{t}{\tau_1}}\right]\widehat{i_\theta}
\end{aligned}
\tag{43}
$$

The charge distribution at the droplet interface is obtained by condition (a3), as:

$$q(t,\theta) = \frac{3\Gamma_R(R-S)}{(R+2)}\left(1 - e^{-\frac{t}{\tau_1}}\right)\cos\theta\tag{44}$$

From the information of Equation 40-Equation 44, we can now deduce the jump in normal and shear stress across the interface, due to (i) electric field and (ii) the electromagnetic coupling. These interfacial stress jumps are responsible for circulation within the droplet and the confined domain, and the deformation of the droplet. The electrical stresses are calculated from the Maxwell stress tensor [13, 53–55]:

$$\tau_{ij}^e = \varepsilon E_i E_j - \frac{1}{2} E_k E_k \varepsilon \delta_{ij} \tag{45}$$

Here, i and j are the tangential and the normal directions, respectively, and $\delta_{ij}$ is the Kronecker delta function.

So, we may express, jump in electrical shear stress is:

$$[\![\tau_{r\theta}^E]\!] = [\![\varepsilon E_r E_\theta]\!] = E_{o,r} E_{o,\theta} - S E_{in,r} E_{in,\theta} = (E_{o,r} - S E_{in,r}) E_{in,\theta} = q E_{in,\theta}$$

$$[\![\tau_{r\theta}^E]\!] = \left[ P_1 + P_2 e^{-\frac{t}{\tau_1}} + P_3 e^{-\frac{2t}{\tau_1}} \right] \sin 2\theta \tag{46}$$

And, jump in normal electrical stress is:

$$[\![\tau_{rr}^E]\!] = \frac{1}{2} [\![\varepsilon \left( E_r^2 - E_\theta^2 \right)]\!] = \frac{1}{2} \left[ \left( E_{o,r}^2 - E_{o,\theta}^2 \right) - S \left( E_{in,r}^2 - E_{in,\theta}^2 \right) \right]$$

$$
\begin{aligned}
[\![\tau_{rr}^E]\!] = & \left[ P_4 + P_5 e^{-\frac{t}{\tau_1}} + P_6 e^{-\frac{2t}{\tau_1}} \right] \cos^2 \theta + \left[ \frac{9}{2} \frac{\Gamma_R^2(S-1)}{(R+2)^2} + \right. \\
& \left. \frac{9\Gamma_R^2 \Gamma_S (S-R) \left( \alpha^3 - 1 \right) (S-1)}{(R+2)^2 (S+2)} e^{-\frac{t}{\tau_1}} + \frac{9\Gamma_R^2 \Gamma_S^2 (S-R)^2 \left( \alpha^3 - 1 \right)^2 (S-1)}{2(R+2)^2 (S+2)^2} e^{-\frac{2t}{\tau_1}} \right]
\end{aligned}
\tag{47}
$$

Where, $[\![\tau_{rr}^E]\!] = \left[ P_4 + P_5 e^{-\frac{t}{\tau_1}} + P_6 e^{-\frac{2t}{\tau_1}} \right] \cos^2\theta +$ Term independent of $\cos^2\theta$

We can express the shear $\left( \tau_{r\theta}^B \right)$ or normal $\left( \tau_{rr}^B \right)$ stress tensor due to the coupling of the electric and magnetic fields as a linear function of the electric and magnetic field, i.e. $\tau_{r\theta}^B$ or $\tau_{rr}^B = f(E, B, \Gamma)$, where $\Gamma$ is the parameter depends on the physical properties of the system. The magnetic effect due to the induced (local) magnetic field is small, so the magnetic field in the magnetic stress tensor is considered to be the applied magnetic field and not the local magnetic field. Since $[\![\tau_{r\theta}^E]\!] = f(\sin 2\theta)$ and $[\![\tau_{rr}^E]\!] = f\left( \cos^2 \theta \right)$, so $[\![\tau_{r\theta}^B]\!]$ and $[\![\tau_{rr}^B]\!]$ necessarily should be the function of $\sin 2\theta$ and $\cos^2\theta$, respectively to satisfy the interfacial condition (b5) and (b6) respectively. Now by similarity and dimensional analysis of $[\![\tau_{r\theta}^B]\!]$ and $[\![\tau_{r\theta}^E]\!]$ of interfacial condition (b5) gives $\Gamma = \sigma l_c \cos\theta$.

So, Jump in shear stress due to electromagnetic coupling is as:

$$[\![\tau_{r\theta}^B]\!] = [\![\sigma E_\theta B \cos\theta]\!] = \left( E_{o,\theta} B \cos\theta - R E_{in,\theta} B \cos\theta \right) = (1 - R) E_{in,\theta} B \cos\theta$$

$$[\![\tau_{r\theta}^B]\!] = -\frac{3}{2} \frac{\Gamma_R(1-R)B}{(R+2)} \left[ 1 - \frac{\Gamma_S(S-R) \left( 1 - \alpha^3 \right)}{(S+2)} e^{-\frac{t}{\tau_1}} \right] \sin 2\theta \tag{48}$$

Jump in normal stress due to electromagnetic coupling is as:

$$[\![\tau_{rr}^B]\!] = [\![\sigma E_r B \cos\theta]\!] = \left( E_{o,r} B \cos\theta - R E_{in,r} B \cos\theta \right) = \left( E_{o,r} - R E_{in,r} \right) B \cos\theta$$

$$[\![\tau_{rr}^B]\!] = \frac{3B\Gamma_r^2 \Gamma_S (S-R) \cos^2\theta}{(S+2)} e^{-\frac{t}{\tau_1}} \tag{49}$$

### C.  Solution for droplet shape deformation parameter (D)

Using Equation 24, Equation 26, Equation 46, Equation 48 with condition (b5), we obtain the relationship

$$
\begin{aligned}
& \frac{1}{2} \left[ (-6A_o - 16B_o - 6C_o - 16D_o) + \lambda \left( 6C_{in} + 16D_{in} \right) \right] + \left[ P_1 + P_2 e^{-\frac{t}{\tau_1}} + P_3 e^{-\frac{2t}{\tau_1}} \right] - \\
& \frac{3}{2} \frac{\Gamma_R(1-R)B}{(R+2)} \left[ 1 - \frac{\Gamma_S(S-R) \left( 1 - \alpha^3 \right)}{(S+2)} e^{-\frac{t}{\tau_1}} \right] = 0
\end{aligned}
\tag{50}
$$

From Equation 49, $B_o$ is obtained in terms of $\frac{d(D)}{dt}$ as:

$$B_o = \frac{P_1 + P_8 e^{-\frac{t}{\tau_1}} + P_3 e^{-\frac{2t}{\tau_1} + \frac{3\Gamma_R B(R-1)}{2(R+2)} - \frac{f_{10}}{Ca}\frac{dD}{dt}}}{f_{11}} \quad (51)$$

We also obtain the expressions of $C_{in}$, $D_{in}$, $A_o$, $C_o$, and $D_o$ in terms of $\frac{d(D)}{dt}$ by substituting $B_o$ from Equation 51 into Equation 24. We use Equation 24, Equation 25, Equation 47, Equation 49, Equation 51 in condition (b6), and retain only terms of coefficient of $cos^2\theta$ (to fulfill the jump condition of normal stresses across the interface). Subsequently, we obtain the expression for the shape deformation parameter ($D$) as:

$$\frac{dD}{dt} + \frac{8}{f_{18}}D = Ca\left[q_1 + q_2 e^{-t/\tau_1} + q_3 e^{-2t/\tau_1}\right] \quad (52)$$

Here, $f_{18}/8$ is another time scale $\tau_2$ ($\alpha$, $\lambda$), which governs the droplet shape relaxation. Here, $q_1$, $q_2$, and $q_3$ are the functions of R, S, $\lambda$, $\alpha$, and B. Their complete expressions are given in the appendix.

The shape of the droplet is undeformed and spherical at t = 0. Imposing the initial condition, we obtain

$$D(t) = D_{SS}\left(1 - Q_1 e^{-\frac{t}{\tau_1}} - Q_2 e^{-\frac{2t}{\tau_1}} - Q_3 e^{-\frac{t}{\tau_2}}\right) \quad (53)$$

The expressions of $Q_1$, $Q_2$, and $Q_3$ of equation (2.38) are described in the appendix. Here, $D_{ss}$ is the steady-state confined droplet deformation parameter, expressed as:

$$D_{ss} = CaQ_1\tau_2 = \frac{CaQ_1 f_{18}}{8} = \frac{9Ca}{16}\frac{\Gamma_R^2}{(R+2)^2}\phi_M \quad (54)$$

$$\phi_M = \left(R^2 - 2S + 1\right) + \frac{3}{5}(R-S)\wp \quad (55)$$

$$\wp = \frac{5}{3}\frac{f_{17}}{f_{11}}\left[1 - \frac{B\left(R^2 + R - 2\right)}{3(R-S)\Gamma_R}\right] \quad (56)$$

Where, $\phi_M$ is the proposed magnetic discriminating function that predicts the droplet's shape at steady-state when both fields are applied and $\wp$ is a correction factor to Taylor's discriminating function. Both $\phi_M$ and $\wp$ are the functions of (R,S,$\lambda$,$\alpha$,B).

For rapid charging of droplets, droplet shape is governed by time scale $\tau_2$. Now, from $D(t)$ and $d(D)/dt$, the constants of stream functions ($C_{in}$,$D_{in}$,$A_o$,$B_o$,$C_o$,$D_o$) thus the generalised streamline equations and velocity fields may be deduced (see Equation 16 to Equation 24).

## V. RESULTS AND DISCUSSION

### A. Variation of steady-state droplet deformation parameter ($D_{ss}$) with $\alpha$ and B

Figure 2 (a and b) show $D_{ss}$ with varying Ca and magnetic field intensities for different $\alpha$ for the two systems described earlier (Table II). Figure 2a shows our theoretical model ($D_{ss}$ for magnetic field B=0 in unconfined domain) is in good agreement with the reports by Taylor [17] and experiments by Salipante and Vlahovska [42]. For R<S system, Taylor predicted deformation in the form of oblate spheroid. In our case, $D_{ss}$ is negative for oblate and positive for prolate spheroid. This is due to the alignment of the droplet dipole in opposite direction of the electric field. For R<S system, the charging response of surrounding fluid $\left(\tau_{E,out} = \frac{\varepsilon_{out}}{\sigma_{out}}\right)$ is faster than droplet $\left(\tau_{E,in} = \frac{\varepsilon_{in}}{\sigma_{oin}}\right)$. Due to the reverse dipole orientation, the net force due to polarization and Coulombic interactions acts inward at the pole. The net force at the equator thereby acts outwards to ensure no normal flow across the interface. Hence, the droplet deforms to oblate.

Further, for different $\alpha$ (at 0.5 and 0.8) and B=0, our analytical expression of $D_{ss}$ is consistent with reports by Mandal et al. [34]. The confinement effects are not prominent at low $\alpha$. As we increase $\alpha$, significant deviation of $D_{ss}$ from Taylor [17]'s unconfined case is observed. At $\alpha$=0.5, the magnitude of $D_{ss}$ is lesser than the unconfined case. Here, the flow field in the droplet and surroundings will be affected due to interaction with the confining wall. So with an increase in $\alpha$, there is reduction in droplet shape relaxation due to an increase in viscous force [34]. This will lead to a reduction of $D_{ss}$ compared to the unconfined domain. The sense of deformation depends on $[\![\sigma_{rr}^H]\!]$ and $[\![\tau_{rr}^E]\!]$. For unconfined domain, the deformation due to $[\![\sigma_{rr}^H]\!]$ depends only on the magnitude of the R-S such that prolate deformation occurs for R>S and vice-versa. For both confined and unconfined domains, the deformation due to $[\![\tau_{rr}^E]\!]$ for the R>S system is prolate (due to alignment of droplet dipole along the applied electric field). For the R<S system, the deformation in both confined and unconfined domain is oblate. For the unconfined R>S system, the contribution of $[\![\tau_{rr}^E]\!]$ is more than $[\![\sigma_{rr}^H]\!]$, and vice-versa.

From condition (b5) it can be deduced that the drop assumes prolate shape for sum of $[\![\sigma_{rr}^H]\!]$ and $[\![\tau_{rr}^E]\!]$ being positive, and oblate shape for negative value. For a confined R > S system, with increase in $\alpha$, $[\![\sigma_{rr}^H]\!]$ is decreasing, and $[\![\tau_{rr}^E]\!]$ will increase. So even though the reduction in $[\![\sigma_{rr}^H]\!]$ may lead to change in oblate, the dominant $[\![\tau_{rr}^E]\!]$ will lead to overall deformation to the prolate. For a confined R<S system, $[\![\tau_{rr}^E]\!]$ will decrease with increase in $\alpha$. However, with increase in $\alpha$, the negative $[\![\sigma_{rr}^H]\!]$ decreases, and eventually the sign changes to positive at a critical $\alpha$. Consequently, shape reversal from oblate ($\alpha$=0.5 Figure 2a) to prolate ($\alpha$=0.8 Figure 2a) is noted. The steady-state variation of $[\![\sigma_{rr}^H]\!]$ and $[\![\tau_{rr}^E]\!]$ with $\alpha$ for B= 0 T, and B = ±1 T, are given in the Table 2 and 3 (supporting information).

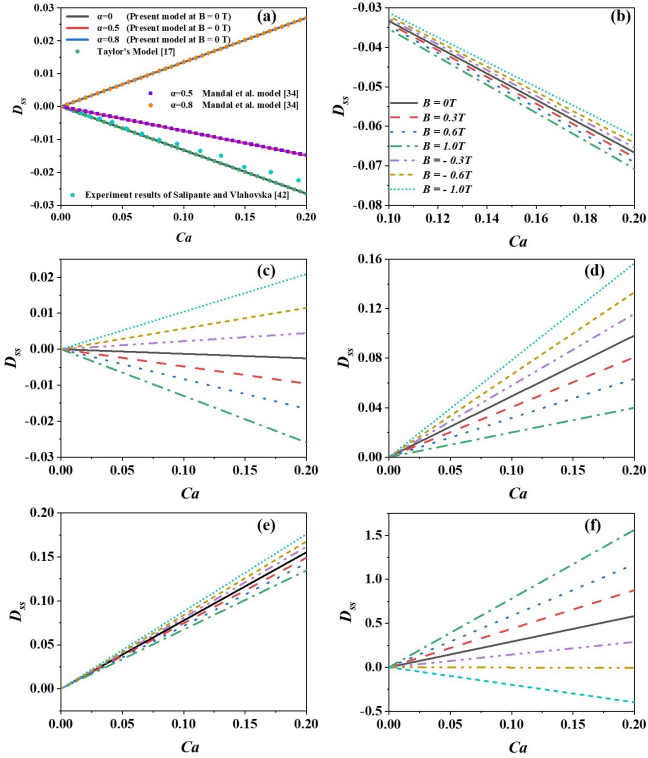

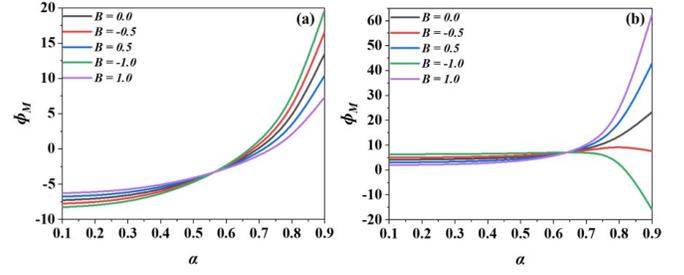

FIG. 3. Variation of $\phi_M$ with $\alpha$ at different B at constant $E_o = 10$ V/mm for (a) System A and (b) System B

FIG. 2. Variation of $D_{ss}$ with Ca for (a) unconfined domain for properties in Ref [42] and B=0, (b) $\alpha = 0.6$, system A, (c) $\alpha = 0.7$, system A, (d) $\alpha = 0.8$, system A, (e) $\alpha = 0.6$, system B, and (f) for $\alpha = 0.9$, system B. Colour and dashed line scheme of figure 2c-2f is same as figure 2b.

Next, we focus on the role of applied magnetic field for different $\alpha$. As the current density at steady state becomes constant and uniform magnetic field is assumed across the interface, $[\![\tau_{rr}^M]\!]$ tends to zero at steady state. Hence, $[\![\tau_{rr}^M]\!]$ will participate in the temporal droplet shape evolution, and near the steady-state it is overshadowed by $[\![\sigma_{rr}^H]\!]$, which dictates the final deformed state along with $[\![\tau_{rr}^E]\!]$. As a result, the magnitude of $[\![\sigma_{rr}^H]\!]$ changes with a change in magnetic field intensity compared to B = 0 T.

Figure 2(b-d) is plotted for the system A. Figure 2b demonstrates that at $\alpha$=0.6, $D_{ss}$ increases with an increase in the B when applied inward (+B), and vice versa for outwards (-B). With the increase of +B, $[\![\sigma_{rr}^H]\!]$ will de-crease compared to B=0, however the sum of stresses is negative (condition b5) and the drop attains oblate shape and $D_{ss}$ will be higher compared to B=0. On the contrary, for B<0, $[\![\sigma_{rr}^H]\!]$ is higher compared to B=0, and thus the $D_{ss}$ is lesser. Up to $\alpha \leq 0.6$, for $-1T \leq B \leq 1T$, the dominance of $[\![\tau_{rr}^E]\!]$ results in oblate spheroid deformation. For $\alpha \geq 0.7$ (Figure 2c), the $D_{ss}$ is reduced for B=0. Now for B>0, as we increase B, the resultant $[\![\sigma_{rr}^H]\!]$ decreases in comparison to B=0. But as the $[\![\tau_{rr}^E]\!]$ is dominant over $[\![\sigma_{rr}^H]\!]$, the net sum of the stresses causes oblate spheroid. Interestingly, shape reversal was observed for B<0, where $[\![\sigma_{rr}^H]\!]$ is increases and dominates over $[\![\tau_{rr}^E]\!]$. Consequently, the $[\![\tau_{rr}^T]\!]$ is positive and causes prolate spheroid.

Similarly, Figure 2(e-f) is plotted for the system B. Along similar lines, Figure 2e demonstrates that for $\alpha$=0.6, the magnitude of $D_{ss}$ increases with an increase in the B when applied inward (-B) and vice versa (+B), exactly in contrast with system A. In such a case, the transition from prolate to oblate is exactly opposite to that of system A. Similar behavior is noted for $\alpha$=0.9 (Figure 2f). Previous studies [17, 34] focused on dynamics of droplets in electric field ambience only predicted that shape will be always prolate for the R>S system. However, we show that on application of the magnetic field in a certain direction, shape reversal is possible due to the dominance of $[\![\sigma_{rr}^H]\!]$ over $[\![\tau_{rr}^E]\!]$.

### B. Variation of $\phi_M$ and $D_{ss}$ with $\alpha$ at different B

It is evident from Equation 54 that $D_{ss}$ is dependent on $\phi_M$. Although the remaining quanties in r.h.s. of Equation 54 are dependent on $\alpha$ and will stay positive. On the contrary, only $\phi_M$ changes sign with change in $\alpha$ and B. Hence, the sign of $D_{ss}$ is solely dependent on $\phi_M$. In this section, we have highlighted the behaviour of $\phi_M$ and $D_{ss}$ with $\alpha$ at different B in Figure 3 and SI.I. (supporting information) respectively for both systems.

$\phi_M$ is dependent on the sum of the magnetic, hydrodynamic and electrical stresses. Similar to the criteria of the sign of the $[\![\tau_{rr}^T]\!]$, $\phi_M$>0 signifies the net effect in the direction of the applied electric field and the steady state shape is a prolate spheroid and vice-versa. Fig-ure 3a is plotted for the R<S system. Negative values of $[\![\tau_{rr}^E]\!]$ implies oblate deformation and the magnitude decreases with increase in $\alpha$. Till $\alpha \leq 0.5$, the magnitude of $[\![\sigma_{rr}^H]\!]$ decreases and the $[\![\tau_{rr}^T]\!]$ remains negative causing oblate droplet deformation. For $0.5 \leq \alpha \leq 0.65$ regime, although the $[\![\sigma_{rr}^H]\!]$ becomes positive and increases with increase in $\alpha$, the $[\![\tau_{rr}^T]\!]$ stays negative results in oblate deformation. On further increase of $\alpha > 0.65$, as $[\![\sigma_{rr}^H]\!]$ dominates over $[\![\tau_{rr}^E]\!]$. The positive $[\![\tau_{rr}^T]\!]$ causes shape reversal from oblate to prolate spheroid. The critical $\alpha$ at which the shape reversal happens varies with B.

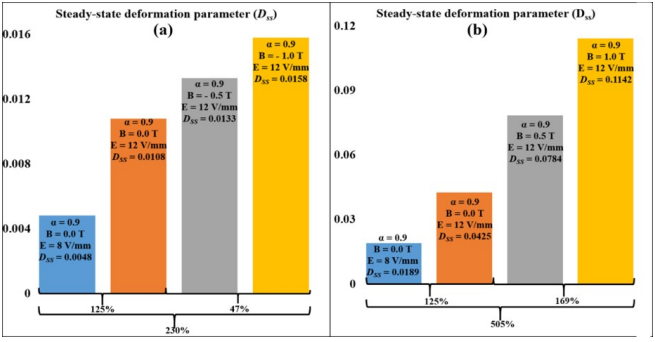

FIG. 4. Variation of $D_{ss}$ with change in electric and magnetic field strength at $\alpha = 0.9$ for (a) System A, (b) System B

Figure 3b illustrates the variation of $\phi_M$ for system B. Upto $\alpha \leq 0.65$, the $[\![\tau_{rr}^E]\!]$ stays positive, increases with increase in $\alpha$. However, the magnitude of $[\![\sigma_{rr}^H]\!]$ decreases results in constant overall sum of the stresses So as evident from Figure 3b, $\phi_M$ assumes a positive constant value and droplet attains prolate spheroid shape. Beyond $\alpha > 0.6$, the $[\![\sigma_{rr}^H]\!]$ changes its sign and magnitude increases with increase in $\alpha$ results in significant change in the $[\![\tau_{rr}^T]\!]$ and consequently in the magnitude of $\phi_M$. Finally, at $\alpha = 0.9$ and B=-1T, the dominance of $[\![\sigma_{rr}^H]\!]$ over $[\![\tau_{rr}^E]\!]$ causes the $[\![\tau_{rr}^T]\!]$ to be of negative. Therefore, the $\phi_M$ assumes a negative value and signifies shape reversal.

## C. Quantitative analysis of $D_{ss}$ with combined EMHD

Figure 4(a and b) show the quantitative analysis of $D_{ss}$ at $\alpha = 0.9$ for both systems. Compared to the base case (B=0T and E=8V/mm), increase of $E_o$ from 8 V/mm to 12 V/mm augments $D_{ss}$ by 125% for both systems. If we proceed from the $2^{nd}$ column (B=0 T) to $4^{th}$ column (B=-1 T for Figure 4a and 1T for Figure 4b), the $D_{ss}$ is increased by $\sim 47\%$ and $\sim 169\%$ for system A and B respectively. So the overall increment in $D_{ss}$ for system A from base case to extreme case ($E_o = 12$ V/mm and B = -1.0 T) is $\sim 230\%$. Therefore it is evident that upon application of magnetic field the increment in $D_{ss}$ is nearly two times the increment of $D_{ss}$ due to presence of $E_0$ only. Similarly for system B, increment in $D_{ss}$ from base case to extreme case ($E_o = 12$ V/mm and B = +1.0 T) is $\sim 505\%$ which is nearly four times the increment of $D_{ss}$ due to only presence of $E_o$. This $\sim 505\%$ of deformation is possible in the absence of a magnetic field when we increase the $E_o$ from 8 V/mm to 16.25 V/mm. So it is evident that if there is a constraint in the magnitude of the $E_o$, the increase in $D_{ss}$ can be achieved by applying magnetic field along with electric field of lower magnitude. Here we have shown quantitative analysis for one extreme condition (i.e. $\alpha$=0.9). This analysis is also valid for all other $\alpha$ also.

## D. Variation of $[\![\tau_{rr}^E]\!]$, $[\![\tau_{rr}^M]\!]$, $[\![\sigma_{rr}^H]\!]$, $[\![\tau_{rr}^T]\!]$ and temporal evolution of $D$ with $\alpha$ at different B

Figure 5(j, k and l) shows the temporal droplet deformation with $\alpha$ for system A at B = -1.0, 0.0 and +1.0T respectively. The variation of $D$ at different $\alpha$ for different B depends on the variation $[\![\tau_{rr}^T]\!]$ shown in Figure 5(g, h and i) for B = -1.0, 0.0 and +1.0T respectively. The variation of $[\![\tau_{rr}^T]\!]$ depends on the individual $[\![\sigma_{rr}^H]\!]$, $[\![\tau_{rr}^E]\!]$ and $[\![\tau_{rr}^M]\!]$ for different B and $\alpha$. The variation of $[\![\tau_{rr}^E]\!]$ for B=0 case is shown in Figure 5b. As $[\![\tau_{rr}^E]\!]$ is independent of B, the temporal variation for non-zero magnetic field cases is same as Figure 5b. The variation of $[\![\tau_{rr}^M]\!]$ is shown in Figure 5(a and c) for B = -1.0 and 1.0T, respectively. The variation of $[\![\sigma_{rr}^H]\!]$ is shown in Figure 5(d, e and f) for B = -1.0, 0 and 1.0T, respectively.

In presence of only electric field, initially, the droplet experiences the polarisation stress due to charge free interface, and attain prolate configuration. Later, due to presence of charges at the interface, shape dynamics are governed by the combined effect of polarisation stress and Coulombic force. The Coulombic force increases over time and tries to deform the droplet in oblate form for R<S system (Figure 5b). For system A, the suspended droplet is less conducting than the surrounding fluid. So, with the increase in $\alpha$, the electrical effect of more conducting fluid decreases and vice-versa, resulting in a reduction of the magnitude of $[\![\tau_{rr}^E]\!]$.

The variations of $[\![\tau_{rr}^M]\!]$ are shown in Figure 5(a and c) at B = -1.0 and +1.0T respectively. When the magnetic field is applied in the -B direction, $[\![\tau_{rr}^M]\!]$ will try for oblate deformation before reaching steady state condition and vice-versa. The magnitude of $[\![\tau_{rr}^M]\!]$ will decrease with time and becomes nearly zero at the steady state.

Now the $[\![\sigma_{rr}^H]\!]$ are affected by both magnetic and electric fields (Figure 5d-f). The $[\![\sigma_{rr}^H]\!]$ depends on $[\![\tau_{rr}^M]\!]$ and $[\![p]\!]$ (i.e. $[\![\sigma_{rr}^H]\!]$=$[\![\tau_{rr}^M]\!]$-$[\![p]\!]$). Figure 5e shows the variation of $[\![\sigma_{rr}^H]\!]$ in the presence of the electric field only. Initially, $[\![\sigma_{rr}^H]\!]$ is due to interaction of conducting fluid and later, the variation is also affected by the interfacial charge interaction with the applied electric field. Initially $[\![\sigma_{rr}^H]\!]$ tends to create prolate deformation. With increase in time, the increase in pressure dominates the overall magnitude of $[\![\sigma_{rr}^H]\!]$ and tends to create oblate deformation. The shape shifting behaviour from prolate to oblate de-formation is observed till $\alpha$=0.567. As we increase $\alpha$, the magnitude of the hydrodynamic stress of the surrounding fluid will decrease as the space between the interface and container surface decreases. The magnitude of the $[\![\sigma_{rr}^H]\!]$ in the region of $0.567 \leq \alpha \leq 0.90$ regimes will try for prolate deformation as the magnitude of the hydrodynamic stress of the suspended droplet will dominate and increase in this region. In the presence of a magnetic field for B = -1.0T, the magnitude of $[\![\sigma_{rr}^H]\!]$ increases with time compared to base case B=0T and vice-versa.

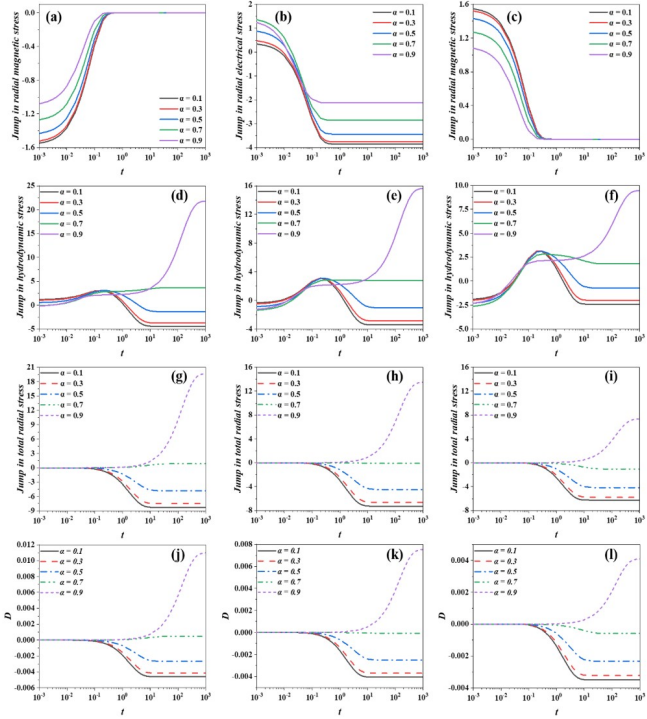

FIG. 5. Transient variation of $[\![\tau_{rr}^M]\!]$ at different $\alpha$ at constant $E_o = 10$ V/mm for system A with (a) B = -1.0 T and (c) B = +1.0T; (b) Variation of $[\![\tau_{rr}^E]\!]$ for B=0T; Variation of $[\![\sigma_{rr}^H]\!]$ with (d) B = -1.0T, (e) B = 0.0T, (f) B = +1.0T; variation of $[\![\tau_{rr}^T]\!]$ with (g) B = -1.0T, (h) B = 0.0T, (i) B = +1.0T. The temporal evolution of $D$ at different $\alpha$ at constant $E_o = 10$ V/mm for system A with (j) B = -1.0T, (k) B = 0.0T, (l) B = +1.0T

The temporal evolution of the droplet depends on the $[\![\tau_{rr}^E]\!]$ across the interface (Figure 5g-i). For base case, polarization and hydrodynamic stress initially act opposite and balance each other, leading to zero $[\![\tau_{rr}^T]\!]$ across the interface, i.e. no deformation (region of $10^{-3} \leq t \leq 10^{-1}$ seconds) in Figure 5(h and k). Later ($t \geq 10^{-1}$ seconds), in the presence of charges at the interface, electrical and hydrodynamic stress act in the same direction, leading to a non-zero $[\![\tau_{rr}^T]\!]$. The $[\![\tau_{rr}^T]\!]$ increases with time until the droplet reaches a steady state condition (region of $10^{-1} \leq t < 10^1$ in Figure 5h). This increase in $[\![\tau_{rr}^T]\!]$ tends to cause oblate deformation. The magnitude of $[\![\tau_{rr}^T]\!]$ varies with the increase in $\alpha$ depends on the variation of $[\![\sigma_{rr}^E]\!]$ and $[\![\sigma_{rr}^H]\!]$. The jump variation of $[\![\tau_{rr}^T]\!]$ for system A decreases with the increase in $\alpha$ as the electrical and hydrodynamic stress decreases with $\alpha$ and the droplet will deform in an oblate shape (Figure 5k). At $\alpha$=0.7, the electrical and hydrodynamic stress will act in opposite directions and balance each other, so there is no droplet deformation. For $\alpha$>0.7, electrical and hydrodynamic stress will still act in opposite directions. However, now the hydrodynamic stress dominates over the electrical stress, so the shape is prolate.

The jump variation of the $[\![\tau_{rr}^T]\!]$ will also depend on the $[\![\tau_{rr}^M]\!]$ ($B \neq 0$) along with the $[\![\tau_{rr}^E]\!]$ and $[\![\sigma_{rr}^H]\!]$ (Figure 5g and i). The jump variation of the $[\![\tau_{rr}^T]\!]$ in the presence of a magnetic field for B =-1.0T increases compared to the base case and vice-versa. This change in the variation of the $[\![\tau_{rr}^T]\!]$ in the presence of the magnetic field is due to the appearance of $[\![\tau_{rr}^M]\!]$ due to electromagnetic coupling and the change in $[\![\sigma_{rr}^H]\!]$ due to interaction of conducting and charged fluid with the electric and magnetic field. At the same time, $[\![\tau_{rr}^E]\!]$ is independent of the B. As the droplet deformation parameter ($D$) of R<S system is dependent on the $[\![\tau_{rr}^T]\!]$ mentioned in interfacial condition (b6), the trends observed in $D$ (Figure 5j-l) are similar to the trends of $[\![\tau_{rr}^T]\!]$ (Figure 5g-i). Figure SI.II (supporting information) shows a similar stress jump and droplet deformation variation analysis for system B. It is noteworthy that for both systems, the deformation increases compared to B=0T when the magnetic field effect act in the direction of the hydrodynamic effect and vice-versa.

## E. Temporal evolution of normalized deformation parameter (D/D$_{ss}$) for different B

Figure 6(a-d) and Figure 6(e-f) highlight the temporal evolution of the $D/D_{ss}$ of system R>S at different B and constant $E_o$=10 V/mm with $\alpha$ varying from 0.2 to 0.8.

At low $\alpha$=0.2 (Figure 6a and e), the deviation of $D/D_{ss}$ from the base case increases with an increase in B. For non-zero B cases, in the +B direction (Figure 6a), the $D/D_{ss}$ curve varies in a non-monotonic way over time. The $D/D_{ss}$ curve shifts in the negative direction with a prominent concave shape in the transient($10^{-2} \leq t \leq 10^1$ secs) region. The degree of concavity increases with the increase in B. In this region, the magnetic field enhances the magnitude of $[\![\sigma_{rr}^H]\!]$. For non-zero B cases in the –B direction (Figure 6e), the $D/D_{ss}$ curve shifts in a positive direction from the base case. The nature of variation of $B \neq 0$ cases is similar to that of B=0T case, i.e. monotonically increasing nature in the transient regime ($3 \times 10^{-3} \leq t \leq 10 secs$) secs. This shifting is due to increase in the magnitude of $[\![\sigma_{rr}^H]\!]$ will lead towards enhancement in $[\![\tau_{rr}^T]\!]$ across the interface compared the base case.

As observed in Figure 6(b-d) and Figure 6(f-h), the degree of deviation from the base curve of $D/D_{ss}$ for B=0T condition decreases with an increase in $\alpha$. As we increase $\alpha$, the $[\![\tau_{rr}^T]\!]$ decreases (refer to the transient variation of stresses in $10^{-2} \leq t \leq 10$ regions in Table 4 and 5 of sup-porting information). For the +B direction at $\alpha \geq 0.8$, the deviation in the transient region is nearly the same as B=0T as the $[\![\tau_{rr}^T]\!]$ across the interface is nearly equal (Figure 6d). In contrast, for the –B direction, even at $\alpha \geq 0.8$, the $[\![\tau_{rr}^T]\!]$ in the transient region is not equal to the same as B=0T. Consequently, the deviation in the transition region is never the same as B=0T (Figure 6h). In the steady state regime, the $[\![\tau_{rr}^M]\!]$ tends to zero due to constant current density across the interface and the assumption of the uniform magnetic field throughout the flow field. So, in the steady state regime, the $D/D_{ss}$ curve converges to a single curve irrespective of the magnitude of B.

For the R<S system (Figure SI.III), the deviation of the $D/D_{ss}$ curve for $B \neq 0$ cases from the base case is not discernible as the magnitude of the $[\![\tau_{rr}^T]\!]$ across the interface is not substantial in transient nature.

## F. Comparative analysis of the evolution of $D$ with EMHD

The influence of electric field intensity on temporal evolution of the $D$ for fixed B and $\alpha$=0.9 are shown in Figure 7(a-c) for R<S system and Figure 7(d-f) for R>S system. The variation of $D$ for other $\alpha$ are similar to the variation of $D$ at $\alpha$=0.9. For variation at other $\alpha$ of both systems, please refer to Figures SI.(IV-VI) of supporting information.

Our analytical model shows that $D$ is dependent on $E_o^2$ (Equation 53). From Figure 7, it is evident that the magnitude of $D$ increases with the increase in $E_o$. Our studies are limited to the small deformation regime. At low $E_o$, the induced magnetic field is negligible and the droplet interface is uniform with no normal flow through it. The increment in the electric field will lead to an increase in the net electrical stress. The increase of the electrical stress is the sole reason for enhanced $D$ for a fixed B. The individual variation of $[\![\tau_{rr}^E]\!]$, $[\![\tau_{rr}^M]\!]$ and $[\![\sigma_{rr}^H]\!]$ with $\alpha$, +B/-B for the different systems are discussed earlier in the preceding section, which determines the deformation dynamics of a droplet. On exceeding a specific limit of $E_o$, large deformation will lead to disintegration of the drop into secondary droplets [45]. The droplet disintegration occurs when the effect of $[\![\tau_{rr}^T]\!]$ increases beyond the limit of the capillary stress.

Figure 7(a-c) shows the variation of $D$ for the R<S system. It is evident that for all the cases presented in Figure 7(a-c) the drop always assumes a prolate shape. In case of B=-1 T, the deformation is more than the base case B=0T as the magnetic field will induce more $\sigma_{rr}^H$ and vice-versa for B=+1T.

Figure 7(d-f) highlights the temporal evolution of $D$ for R>S system. For the +B direction (B=1T), deformation increases compared to B=0T as the magnetic field will increase the $\sigma_{rr}^H$ (Figure 7f). For both B=0 and +1T, the droplet tends to deform in prolate shape. However in case of B=-1T, the drop assumes oblate shape (Fig-

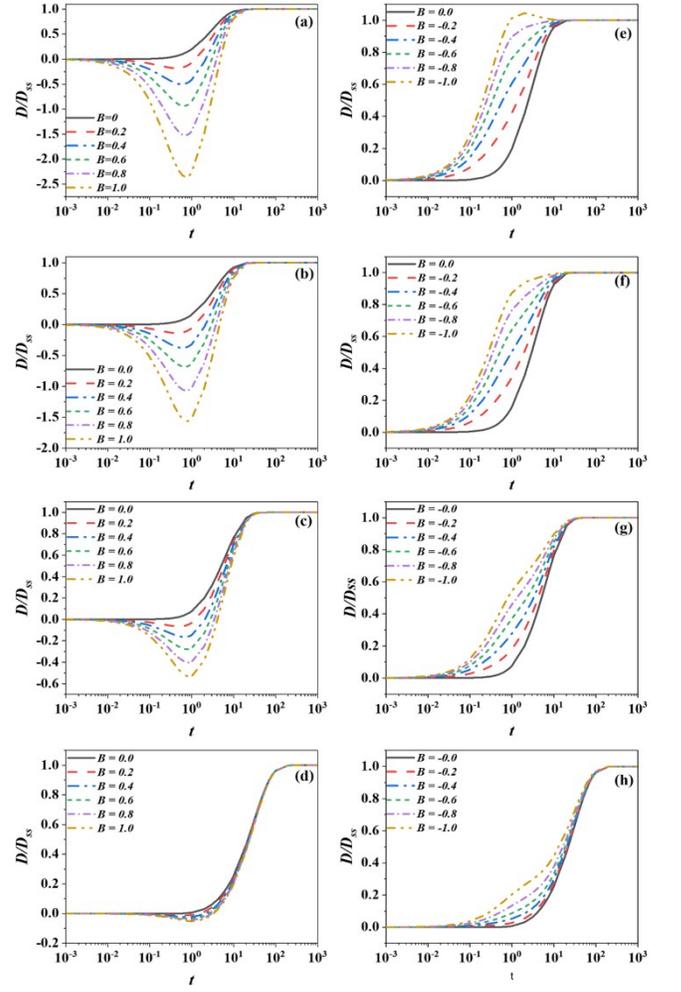

FIG. 6. Transient evolution of $D/D_{ss}$ for different magnetic field strength for system B at constant $E_o = 10$ V/mm i.e. (a) +B, $\alpha = 0.2$, (b) +B, $\alpha = 0.4$, (c) +B, $\alpha = 0.6$, (d) +B, $\alpha = 0.8$, (e) -B, $\alpha = 0.2$, (f) -B, $\alpha = 0.4$, (g) -B, $\alpha = 0.6$, (h) -B, $\alpha = 0.8$

ure 7d). The shape reversal at B=-1T is due to the dominance of the $\sigma_{rr}^H$. In this case, $\sigma_{rr}^H$ acts opposite to the hydrodynamic effect at B=0T resulting in shape shifting from prolate to oblate configuration.

If there is constraint in applying higher electric field strength, we can enhance the deformation of the droplet by modulating the magnitude and direction of B at a lower electric field intensity. Figure 8 shows the effect of B on the temporal evolution of $D$ for different $\alpha$ i.e. 0, 0.5 and 0.9. The electric field intensity is fixed at $E_o = 10$ V/mm for both the systems. For variation at intermediate $\alpha$ for both systems, please refer to Figure SI.VII of the supporting information.

The droplet deformation parameter ($D$) at constant $E_o$ increases or decreases at different B depending on whether the magnetic field enhances or decreases the $\sigma_{rr}^H$. When the magnetic field enhances $\sigma_{rr}^H$, deformation will increase and vice-versa.

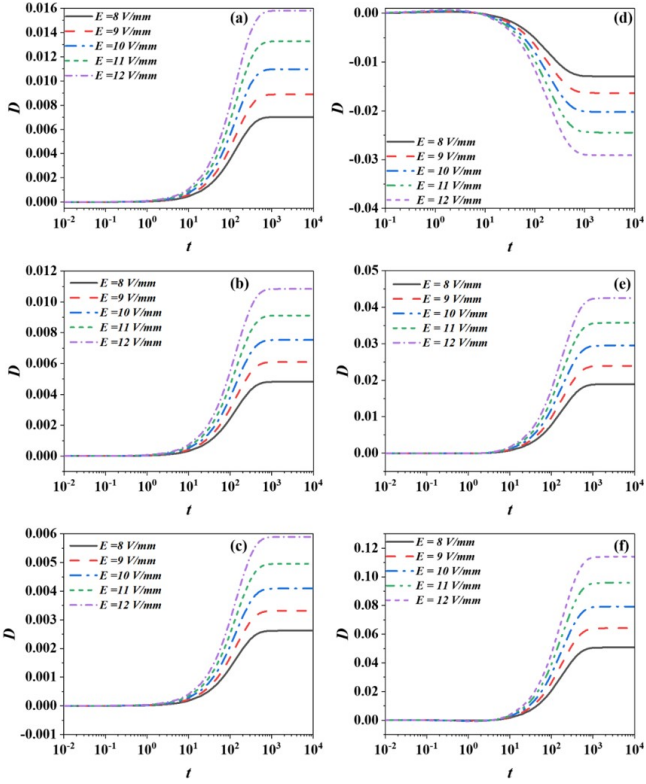

FIG. 7. Temporal deformation of D D for different $E_o$ with $\alpha$ = 0.9 for system A with (a) B = -1.0 T, (b) B = 0.0 T, (c) B = +1.0 T and for system B with (d) B = -1.0 T, (e) B = 0.0 T, (f) B = +1.0 T

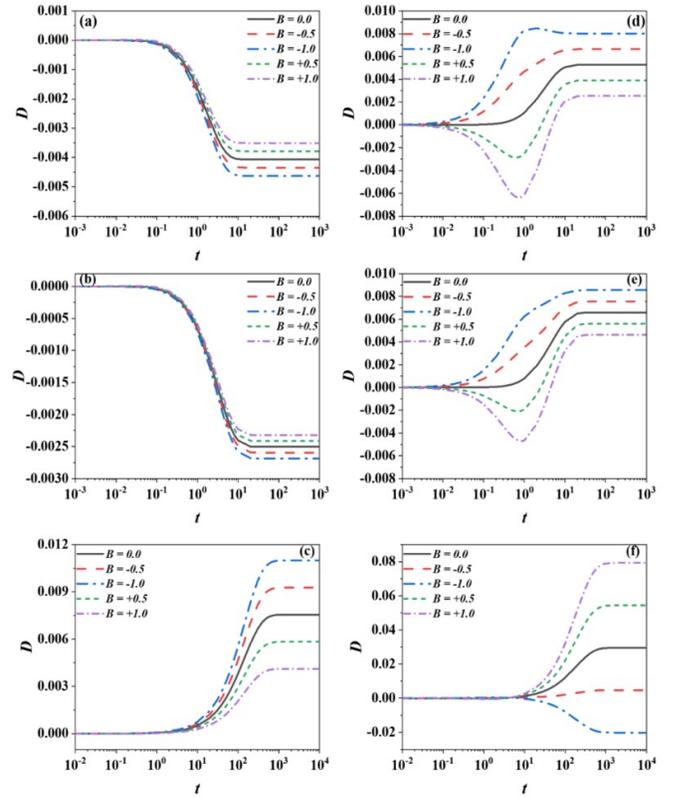

FIG. 8. Temporal deformation of the D for different B at constant $E_o$ = 10 V/mm for system A with (a) $\alpha = 0$, (b) $\alpha$ = 0.5, (c) $\alpha$ = 0.9 and for system B with (d) $\alpha = 0$, (e) $\alpha$ = 0.5, (f) $\alpha$ = 0.9.

[31] showed in their analysis that a drop size of 10 mm deforms in the order of $10^{-3}$ for the applied electric field of 10 V/mm for an unconfined domain. In our case, we are also considering a 10 mm droplet size under 10 V/mm magnitudes of the electric field. The deformation based on present model at B = 0.0 T is in the order of $10^{-3}$ for both the $R < S$ system (Figure 5k, Figure 8a) and for the R>S system (Supplementary Figure SI.II, Figure 8d).

For R<S system (Figure 8a and b), although the magnitude of $D$ is changing with change in direction and magnitude of B, the overall configuration is of oblate shape identical to the base case. For the highest $\alpha = 0.9$, shape reversal from oblate to prolate configuration is observed for all B's. In this case the influence of B on $\sigma_{rr}^H$ is more than that of electric field. In addition, the resultant $\sigma_{rr}^H$ acts in opposite direction of the electric field, thereby causing the shape reversal.

For R>S system, from Figure 8d-e it is evident that although the drop tends to achieve oblate shape (B=+0.5 and +1T) in the transient regime ($10^{-2} \leq t \leq 10$ secs), the steady state droplet configuration for all B's is prolate shape. However for the highest $\alpha = 0.9$, even the steady state configuration for B=-1 T is oblate, which is different from the base case.

## G. Distribution of the internal and external EMHD flow field

The mutual interaction of the electric and magnetic fields generate some unique patterns of streamlines within the droplet and the surrounding fluid. Figure 9 and Figure 10 show the temporal evolution of streamlined patterns for system A and B with different B at $\alpha$=0.8 (refer Figure SI.VIII to Figure SI.XV of supporting information for other $\alpha$) and $E_o$=10V/mm.

Since the applied electric field is of irrotational type, the induced magnetic field is negligible. The charge relaxation time is not affected much by the externally applied magnetic field. All the charges accumulate in the interface ($\sim \leq 10^{-1}$sec) after applying the electric field. So the variation of electric potential, electric field, electrical stress and $[\![\tau_{rr}^E]\!]$ across the interface are not affected much by the externally applied magnetic field. As the charge accumulation on the interface takes place, the coupled effect of the magnetic field and the electric field becomes significant. The combined effect of $[\![\tau_{r\theta}^E]\!]$ and $[\![\tau_{r\theta}^M]\!]$ will influence the flow field within the droplet and the surrounding medium. In comparison to B=0 T case, the streamline density will increase on application of both electric and magnetic fields. This streamlined density

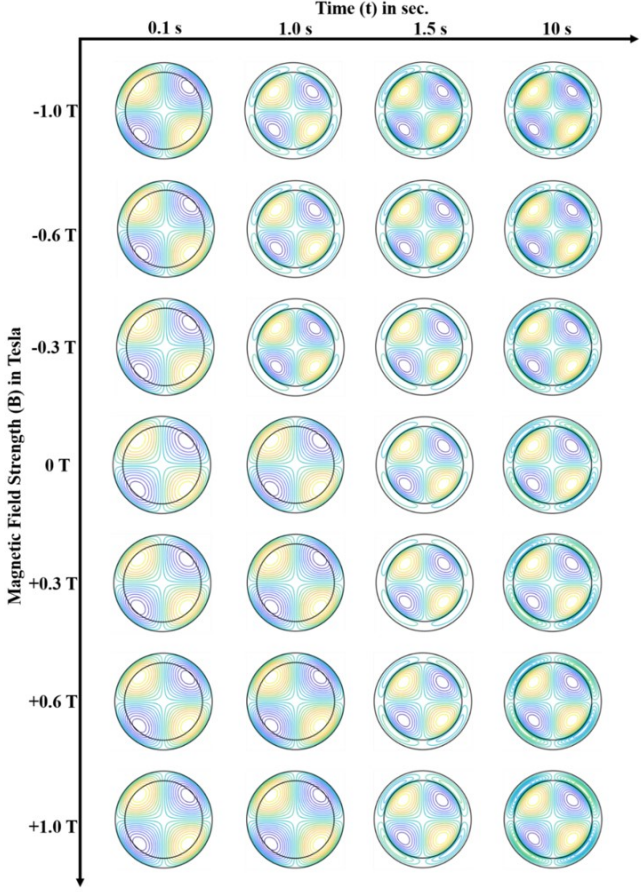

FIG. 9. Transient evolution of streamline pattern for different $E_o = 10$ V/mm at $\alpha = 0.8$ for System A. The four vortices inside the liquid droplet match their ambient fluid counterparts. The yellow and blue colour structure shows the positive and negative values of stream functions, respectively, at corresponding r and $\theta$ values. This colour scheme is applicable for all contours.

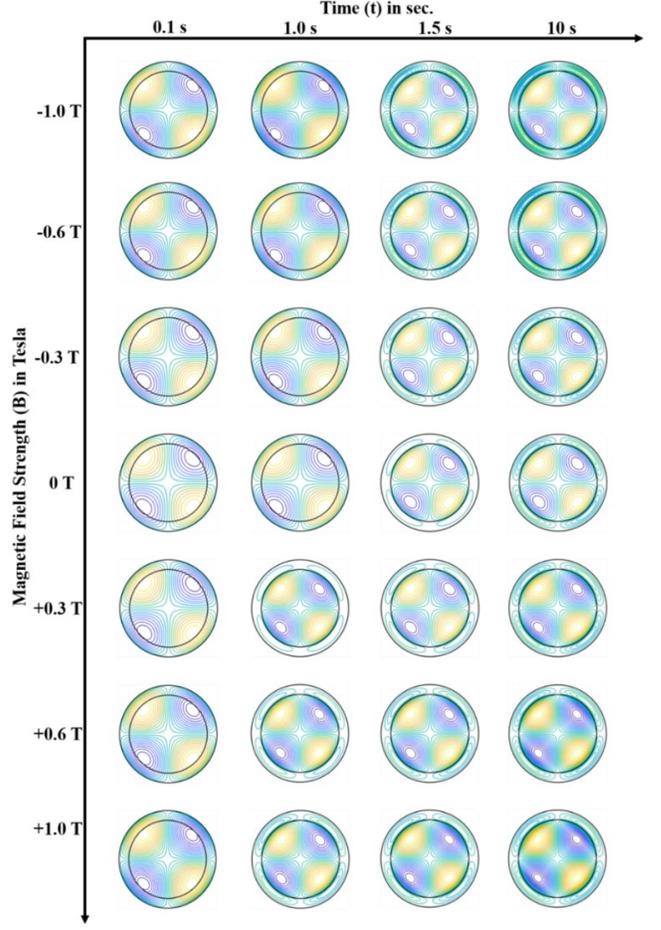

FIG. 10. Transient evolution of streamline pattern for different B at constant $E_o = 10$ V/mm at $\alpha = 0.8$ for System B

depends on $E_o$, B, properties of the system, and $\alpha$. The magnitude of the $[\![\tau_{r\theta}^E]\!]$ is unaffected by the presence of a magnetic field. Without a magnetic field, interfacial $[\![\tau_{r\theta}^E]\!]$ is balanced by $[\![\tau_{r\theta}^H]\!]$. But in the presence of an external magnetic field, $[\![\tau_{r\theta}^E]\!]$ is balanced by both $[\![\tau_{r\theta}^H]\!]$ and $[\![\tau_{r\theta}^M]\!]$, dictated by the interfacial conditions (b5). As $[\![\tau_{r\theta}^E]\!]$ is invariant of B, the variation in $[\![\tau_{r\theta}^H]\!]$ is in synchrony with the variation in $[\![\tau_{r\theta}^M]\!]$ (refer interfacial condition b5). In a previous study by Mandal *et al.* [34], they reported the presence of crowded and closed streamlines of highly confined droplets in presence of electric field only i.e. B=0T. In presence of magnetic field, the $[\![\tau_{r\theta}^H]\!]$ increases more across the interface in comparison to B=0T case. Consequently, denser streamlines are observed either in the droplet or within the confined space. For R<S system and +B direction, the streamlines will be denser in the surrounding medium (Figure 9). For –B direction, the streamlines will be denser in the confined drop and vice-versa for R>S systems for +B and

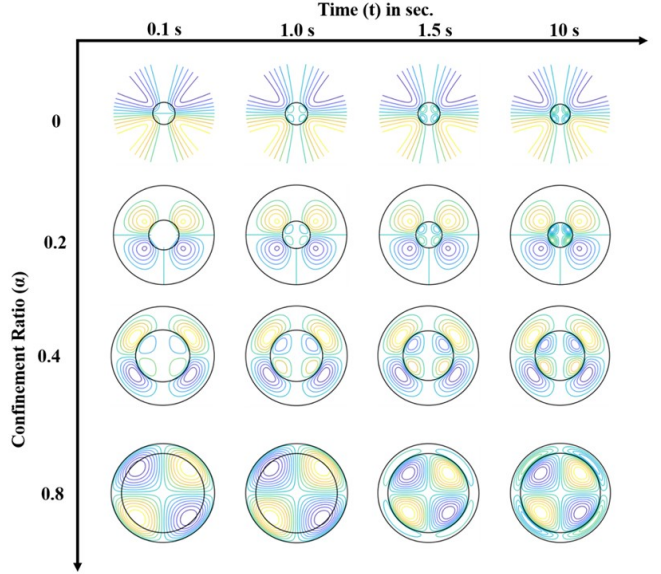

FIG. 11. Transient evolution of streamline pattern for different $\alpha$ at constant $E_o = 10$ V/mm at B = 0.0 T for system B

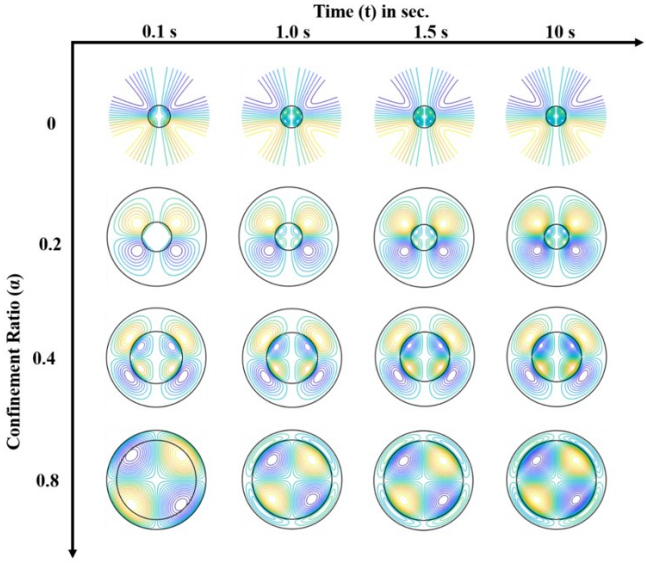

FIG. 12. Transient evolution of streamline pattern for $\alpha$ at constant $E_o = 10$ V/mm at B = +1.0 T for system B

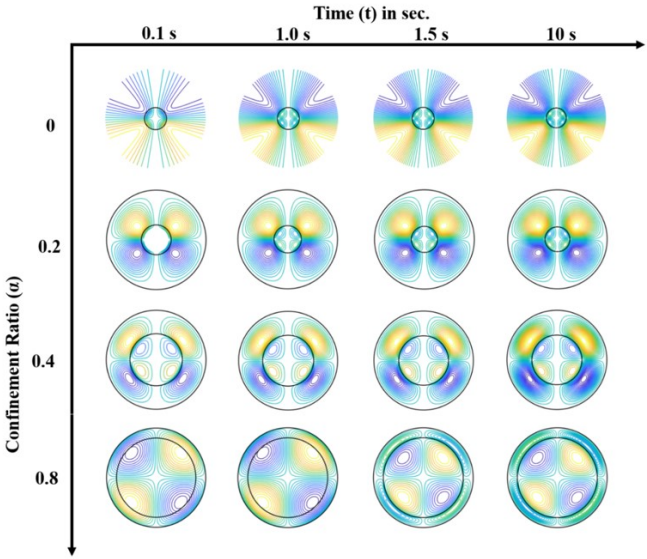

FIG. 13. Transient evolution of streamline pattern for different $\alpha$ at constant $E_o = 10$ V/mm at B = -1.0 T for system B

–B directions (Figure 10).

Along each row the temporal evolution of streamlines are highlighted (Figure 9 and Figure 10). Initially, the streamlines will appear as closed loops present throughout both the domains of the drop and the surrounding fluid (1$^{st}$ column of Figure 9 and Figure 10). As time progresses, more charges accumulate on the interface and will enhance the effect of electrical and magnetic fields. As a result the $\tau_{r\theta}^H$ is also enhanced. Due to the increase in the magnitude of $\tau_{r\theta}^H$ with time, the earlier observed closed looped streamlines will be no longer possible. From time $\sim 1$ sec onwards, due to the stress mismatch across the interface, the earlier observed closed loop streamlines (1$^{st}$ column Figure 9, Figure 10) will evolve into separate streamline contours in the two flow domains. With increase in time, further increment in the jump of $\tau_{r\theta}^H$ will increase the density of the contours either in droplets or in the surrounding medium.

Now, consider the streamline variation in R<S system with $\alpha$=0.8 and B = +1.0 T (last row of Figure 9). For this condition, $\tau_{r\theta}^M$ in the droplet is more than $\tau_{r\theta}^M$ in the surrounding medium. As the $\tau_{r\theta}^H$ variation is opposite to the variation in $\tau_{r\theta}^M$ due to interfacial condition b5, the $\tau_{r\theta}^H$ in drop is lesser than that of the surrounding medium. Therefore, on application of magnetic field, $\tau_{r\theta}^H$ in confined space increases further in comparison to B=0T case. Hence, the surrounding medium has presence of denser streamlines than the droplet. Similar trends are observed in +0.3 T and 0.6 T cases (5$^{th}$ - 6$^{th}$ row of Figure 9). For the same $\alpha$=0.8 and R<S system, variation in $\tau_{r\theta}^M$ and $\tau_{r\theta}^H$ across the interface will be in opposite direction in case of –B direction. As a result, denser streamlines are observed within the droplet (1$^{st}$ - 3$^{rd}$ row of Figure 9). For both +B and –B directions of magnetic field, with increase in B, magnitude of $[\![\tau_{r\theta}^M]\!]$ increases. So for both +B and –B directions, the density of streamlines increases in the surrounding medium (5$^{th}$ - 7$^{th}$ row of Figure 9) and the confined drop (1$^{st}$ - 3$^{rd}$ row of Figure 9) respectively.

The streamline variation in the R>S system in the presence of the magnetic field strength with $\alpha$=0.8 is shown in Figure 10. Here, the droplet is more conducting than the surrounding fluid. Since the effect of the magnetic field will be more dominant in the R>S system than the R<S system, denser streamlines are observed within the droplet and the surrounding medium. The variation in $\tau_{r\theta}^M$ and $\tau_{r\theta}^H$ across the interface for the R>S system will be in the opposite direction compared to the R<S system. As a result, for the +B direction, the droplet has a presence of denser streamlines than the surrounding medium (4$^{th}$ - 6$^{th}$ row of Figure 10). For the –B direction, denser streamlines are observed in the surrounding medium (1$^{st}$ - 3$^{rd}$ row of Figure 10).

Figure 11 highlights the role of $\alpha$ on the temporal evolutions of streamlined patterns (for R>S system) at B=0T. The temporal evolution of the streamlines are in good agreement with the previous study by Mandal et al. [34]. For unconfined domain, the streamline contours are confined to individual domains (1$^{st}$ row of Figure 11) unlike the confined domains (Figure 9 and Figure 10). With the increase in $\alpha$, the contours are more stretched in the surrounding medium along the curvature of the droplet. This implies the existence of strong velocity gradients along the curvature (tangential). The contours within the droplet also enhance with the increase in $\alpha$ due to the increase of velocity gradient along the radial direction.

Figure 12 and Figure 13 show the temporal evolutions of streamlined patterns (for R>S system) at different $\alpha$ for B = +1.0 and -1.0T, respectively. In the presence of the magnetic field, the streamline contours are denser than the B=0T case. The direction of the magnetic field strength dictates the relative density of streamline contours within the droplet and in the surrounding medium, i.e. denser streamlines within the droplet for the +B direction (Figure 12 for B = +1.0T) and in the surrounding fluid for the −B direction (Figure 13 for B = -1.0T). The streamline contours for the intermediate B, i.e. B = ±0.3T and B = ±0.6T for systems A and B, are shown from Figure SI.XVI to Figure SI.XXIII of supporting section. In these intermediate magnetic field strengths, the streamline contours are lesser denser than B = ±1.0T and denser than B=0.0T for both systems.

## VI. CONCLUSIONS

The present analytical study focuses on the deformation dynamics and the flow field analysis of a droplet suspended in a confined fluid in the presence of both electric and magnetic fields. Our analysis proposed an expression for the droplet deformation parameter ($D$) varies exponentially with time for both instantaneous (System A) and finite charge relaxation time (System B). The droplet deformation parameter ($D$) depends on the $E_o$, the strength and the direction of the applied B, $\alpha$ and properties of droplet and surrounding fluid. We also developed a magnetic discriminating function ($\phi_M$) which predicts the steady-state shape of the droplet when both electric and magnetic fields are present. $\phi_M<0$ implies the evolution of the droplet from spherical shape to oblate spheroid while >0 corresponds to prolate spheroid and no deformation for $\phi_M=0$. After the initial validation of confined drop deformation in presence of electric field only (B=0T) with previous studies [13, 34], we explored the contribution of $[\![\sigma_{rr}^H]\!]$, $[\![\tau_{rr}^E]\!]$ and $[\![\tau_{rr}^M]\!]$ in transient deformation and flow field analysis for $B \neq 0$T cases.

We have shown that the deformation dynamics is governed by the $[\![\tau_{rr}^T]\!]$. With an increase in $\alpha$, the magnitude of $[\![\tau_{rr}^E]\!]$ decreases and increases for the R<S and R>S system, respectively. The $[\![\tau_{rr}^M]\!]$ will increase with the increase in B. The $[\![\sigma_{rr}^H]\!]$ is governed by both electric and magnetic field. When $[\![\sigma_{rr}^E]\!]$ and $[\![\sigma_{rr}^M]\!]$ tend to deform the droplet in the same direction, the droplet deformation will increase and vice-versa. In specific cases, when both $[\![\sigma_{rr}^H]\!]$ and $[\![\tau_{rr}^E]\!]$ tend to deform the droplet in the opposite direction, shape reversal will be observed when $[\![\sigma_{rr}^H]\!]$ dominates over $[\![\tau_{rr}^E]\!]$.

The flow field analysis of the present work uncovers the effect of the magnetic field on $[\![\tau_{r\theta}^H]\!]$. The streamlined density within the droplet and in the surrounding fluid enhances with the increase in magnetic field intensity.

For the R<S system, streamline density is more within the droplet in the −B direction and the surrounding fluid in the +B direction and vice-versa for the R>S system. Our analysis also reveals that in the presence of the magnetic field, for the R>1 system, the droplet deformation and flow field intensity is more than in the R<1 system. As the droplet is more conducting than the surrounding fluid for the R>1 system, higher $[\![\sigma_{rr}^M]\!]$ will occur due to electric and magnetic fields. The present analysis shows that even at lower electric field intensity, we can augment the deformation dynamics by incorporating magnetic field. The manipulation of droplet deformation and flow field in a specific direction by modulating the magnetic field with an electric field may have potential use in various microfluidic applications.

## DATA AVAILABILITY STATEMENT:

All data have been included in the main paper and supporting information document.

## DECLARATION OF INTERESTS:

The authors report no conflicts of interest with respect to this research.

## ACKNOWLEDGEMENTS:

The authors thank IIT Ropar and IIT Kharagpur for supporting the work through internal funds.

## Appendix

Equation 23 contains the functions $f_1 - f_5$ of the following form:

$$
\begin{aligned}
f_1 &= 2 - 5\alpha^3 + 3\alpha^5, \\
f_2 &= 2 - 7\alpha^5 + 5\alpha^7, \\
f_3 &= 5 - 7\alpha^2 + 2\alpha^7, \\
f_4 &= 3 - 5\alpha^2 + 2\alpha^5, \\
f_5 &= -4 + 25\alpha^3 - 42\alpha^5 - 4\alpha^{10} + 25\alpha^7
\end{aligned}
\tag{A.1}
$$

Equation 46 contains the functions $P_1 - P_3$ & Equation 47 contains the functions $P_4 - P_6$ of the following form:

$$P_1 = \frac{9\Gamma_R^2\left(\,S-R\right)}{2(R+2)^2}$$

$$P_2 = -P_1\left[1+\frac{\Gamma_S(S-R)\left(1-\alpha^3\right)}{(S+2)}\right]$$

$$P_3 = P_1\left[\frac{\Gamma_S(S-R)\left(1-\alpha^3\right)}{(S+2)}\right]$$

$$P_4 = \frac{9\Gamma_R^2\left(R^2-2\,S+1\right)}{2(R+2)^2}$$

$$P_5 = \frac{9\Gamma_R^2\Gamma_S(S-R)}{2(R+2)^2\left(\,S+2\right)}\left[2R\left(2+\alpha^3\right)+2\left(\alpha^3-1\right)+4\,S\left(1-\alpha^3\right)^2\right]$$

$$P_6 = \frac{9\Gamma_R^2\Gamma_S^2\left(\,S-R\right)^2}{2(R+2)^2\left(\,S+2\right)^2}\left[\left(2+\alpha^3\right)^2+\left(\alpha^3-1\right)^2-2\,S\left(1-\alpha^3\right)^2\right]$$

(A.2)

Equation 53 contains the functions $q_1 - q_3$ of the following form:

$$q_1 = \frac{1}{f_{18}}\left(P_4-\frac{P_1 f_{17}}{f_{11}}-\frac{3\Gamma_R B}{2}\frac{\left(R-1\right)f_{17}}{(R+2)\,f_{11}}\right)$$

$$q_2 = \frac{1}{f_{18}}\left(P_{10}-\frac{P_8 f_{17}}{f_{11}}\right), q_3 = \frac{1}{f_{18}}\left(P_6-\frac{P_3 f_{17}}{f_{11}}\right)$$

(A.3)

Equation 54 contains the functions $Q_1 - Q_3$ of the following form:

$$Q_1 = \frac{q_2\tau_3}{q_1\tau_2},\quad Q_2 = \frac{q_3\tau_4}{q_1\tau_2},\quad Q_3 = \frac{q_1\tau_2-q_2\tau_3-q_3\tau_4}{q_1\tau_2}$$

(A.4)

Equation 56 contains the functions $P_8$ of the following form:

$$P_7 = \frac{3}{2}\Gamma_R\Gamma_S B\left(1-\alpha^3\right)\left[\frac{(R-1)(S-R)}{(R+2)(S+2)}\right],\quad P_8 = P_2-P_7$$

(A.5)

$q_2$ contains the functions $P_{10}$ of the following form:

$$P_9 = \frac{3B\Gamma_R^2\Gamma_S(S-R)}{(S+2)},\quad P_{10} = P_5+P_9$$

(A.6)

Expression of the functions $f_6 - f_{18}$ is of the following form:

$$
\begin{aligned}
&f_6 = \tfrac{16\alpha^5-10\alpha^3+4}{f_1}, &&f_7 = \tfrac{8f_1-3f_2+3\alpha^3f_3-8\alpha^5f_4}{f_1}, &&f_8 = \tfrac{16\alpha^5-10\alpha^3-6}{f_1},\\
&f_9 = \tfrac{5f_5}{2f_1}, &&f_{10} = f_6-\lambda f_8, &&f_{11} = f_7-\lambda f_9,\\
&f_{12} = \tfrac{6\alpha^5+20\alpha^3+24}{f_1}, &&f_{13} = \tfrac{24f_1-18f_2-6\alpha^3f_3-3\alpha^5f_4}{f_1}, &&f_{14} = \tfrac{-6\alpha^5-20\alpha^3+26}{f_1},\\
&f_{15} = -\tfrac{9f_5}{2f_1}, &&f_{16} = f_{12}+\lambda f_{14}, &&f_{17} = f_{13}+\lambda f_{15},
\end{aligned}
$$

(A.7)

[1] W. A. Macky, Some investigations on the deformation and breaking of water drops in strong electric fields, Proceedings of the Royal Society of London. Series A, Containing Papers of a Mathematical and Physical Character **133**, 565 (1931).

[2] Y. Wu and R. L. Clark, Electrohydrodynamic atomization: A versatile process for preparing materials for biomedical applications, Journal of Biomaterials Science, Polymer Edition **19**, 573 (2008).

[3] A. Banerjee, E. Kreit, Y. Liu, J. Heikenfeld, and I. Papautsky, Reconfigurable virtual electrowetting channels, Lab on a Chip **12**, 758 (2012).

[4] K. Ikemoto, I. Sakata, and T. Sakai, Collision of millimetre droplets induces DNA and protein transfection into cells, Scientific Reports **2**, 1 (2012).

[5] M. Abdelaal and M. Jog, Steady and time-periodic electric field-driven enhancement of heat or mass transfer to a drop: Internal problem, International Journal of Heat

and Mass Transfer **55**, 251 (2012).

[6] O. A. Basaran, Small-scale free surface flows with breakup: Drop formation and emerging applications, AIChE Journal **48**, 1842 (2002).

[7] C. Xiaopeng, C. Jiusheng, and Y. Xiezhen, Advances and applications of electrohydrodynamics, Chinese Science Bulletin **48**, 1055 (2003).

[8] K. J. Ptasinski and P. J. Kerkhof, Electric Field Driven Separations: Phenomena and Applications, Separation Science and Technology **27**, 995 (1992).

[9] U. Ghosh, A consistent description of electro-magneto-hydrodynamic flows in narrow slits, Journal of Physics D: Applied Physics **53**, 10.1088/1361-6463/ab9049 (2020).

[10] D. Si and Y. Jian, Electromagnetohydrodynamic (EMHD) micropump of Jeffrey fluids through two parallel microchannels with corrugated walls, Journal of Physics D: Applied Physics **48**, 10.1088/0022-3727/48/8/085501 (2015).

[11] A. Sinha and G. C. Shit, Electromagnetohydrodynamic flow of blood and heat transfer in a capillary with thermal radiation, Journal of Magnetism and Magnetic Materials **378**, 143 (2015).

[12] A. Esmaeeli and P. Sharifi, The transient dynamics of a liquid column in a uniform transverse electric field of small strength, Journal of Electrostatics **69**, 504 (2011).

[13] A. Esmaeeli and A. Behjatian, Electrohydrodynamics of a liquid drop in confined domains, Physical Review E **86**, 036310 (2012).

[14] R. J. Haywood, M. Renksizbulut, and G. D. Raithby, Transient deformation of freely-suspended liquid droplets in electrostatic fields, AIChE Journal **37**, 1305 (1991).

[15] O. Vizika and D. A. Saville, The electrohydrodynamic deformation of drops suspended in liquids in steady and oscillatory electric fields, Journal of Fluid Mechanics **239**, 1 (1992).

[16] J. Zhang, J. D. Zahn, and H. Lin, Transient solution for droplet deformation under electric fields, Physical Review E - Statistical, Nonlinear, and Soft Matter Physics **87**, 1 (2013).

[17] G. Taylor, Studies in electrohydrodynamics. I. The circulation produced in a drop by an electric field, Proceedings of the Royal Society of London. Series A. Mathematical and Physical Sciences **291**, 159 (1966).

[18] G. Taylor, Disintegration of water drops in an electric field, Proceedings of the Royal Society of London. Series A. Mathematical and Physical Sciences **280**, 383 (1964).

[19] R. S. Allan and S. G. Mason, Particle behaviour in shear and electric fields I. Deformation and burst of fluid drops, Proceedings of the Royal Society of London. Series A. Mathematical and Physical Sciences **267**, 45 (1962).

[20] J. R. Melcher and G. I. Taylor, Electrohydrodynamics: A Review of the Role of Interfacial Shear Stresses, Annual Review of Fluid Mechanics **1**, 111 (1969).

[21] D. A. Saville, Electrohydrodynamics: The Taylor-Melcher Leaky Dielectric Model, Annual Review of Fluid Mechanics **29**, 27 (1997).

[22] J. Hua, L. K. Lim, and C.-H. Wang, Numerical simulation of deformation/motion of a drop suspended in viscous liquids under influence of steady electric fields, Physics of Fluids **20**, 10.1063/1.3021065 (2008).

[23] C. Liu, D. Springer, and G. D. Clifford, A consistent description of electro-magneto-hydrodynamic flows in narrow slits, Physiological Measurement **38**, 1730 (2017).

[24] G. I. Taylor, The formation of emulsions in definable fields of flow, Proceedings of the Royal Society of London. Series A, Containing Papers of a Mathematical and Physical Character **146**, 501 (1934).

[25] Rosenkilde, C.E., Dielectric Fluid Drop in an Electric Field, Proceedings of the Royal Society of London. Series A. Mathematical and Physical Sciences **312**, 473 (1969).

[26] J. Q. Feng and T. C. Scott, A computational analysis of electrohydrodynamics of a leaky dielectric drop in an electric field, Journal of Fluid Mechanics **311**, 289 (1996).

[27] J.-W. Ha and S.-M. Yang, Deformation and breakup of Newtonian and non-Newtonian conducting drops in an electric field, Journal of Fluid Mechanics **405**, 131 (2000).

[28] E. Lac and G. M. Homsy, Axisymmetric deformation and stability of a viscous drop in a steady electric field, Journal of Fluid Mechanics **590**, 239 (2007).

[29] Y. Lin, P. Skjetne, and A. Carlson, A phase field model for multiphase electro-hydrodynamic flow, International Journal of Multiphase Flow **45**, 1 (2012).

[30] J. M. López-Herrera, S. Popinet, and M. A. Herrada, A charge-conservative approach for simulating electrohydrodynamic two-phase flows using volume-of-fluid (2011).

[31] A. Esmaeeli and P. Sharifi, Transient electrohydrodynamics of a liquid drop, Physical Review E - Statistical, Nonlinear, and Soft Matter Physics **84**, 1 (2011).

[32] C. Sozou, Electrohydrodynamics of a liquid drop: the development of the flow field, Proceedings of the Royal Society of London. A. Mathematical and Physical Sciences **334**, 343 (1973).

[33] J. A. Lanauze, L. M. Walker, and A. S. Khair, The influence of inertia and charge relaxation on electrohydrodynamic drop deformation, Physics of Fluids **25**, 10.1063/1.4826609 (2013).

[34] S. Mandal, K. Chaudhury, and S. Chakraborty, Transient dynamics of confined liquid drops in a uniform electric field, Physical Review E - Statistical, Nonlinear, and Soft Matter Physics **89**, 10.1103/PhysRevE.89.053020 (2014).

[35] S. Moriya, K. Adachi, and T. Kotaka, Deformation of Droplets Suspended in Viscous Media in an Electric Field. 1. Rate of Deformation, Langmuir **2**, 155 (1986).

[36] T. Nishiwaki, K. Adachi, and T. Kotaka, Deformation of Viscous Droplets in an Electric Field: Poly(propylene oxide)/Poly(dimethylsiloxane) Systems, Langmuir **4**, 170 (1988).

[37] T. Tsukada, T. Katayama, Y. Ito, and M. Hozawa, Theoretical and Experimental Studies of Circulations Inside and Outside a Deformed Drop Under a Uniform Electric Field (1993).

[38] J. Q. Feng, Electrohydrodynamic behaviour of a drop subjected to a steady uniform electric field at finite electric Reynolds number, Proceedings of the Royal Society A: Mathematical, Physical and Engineering Sciences **455**, 2245 (1999).

[39] C. T. Wilson and G. I. Taylor, The bursting of soap-bubbles in a uniform electric field, Mathematical Proceedings of the Cambridge Philosophical Society **22**, 728 (1925).

[40] J. D. Sherwood, Breakup of fluid droplets in electric and magnetic fields, Journal of Fluid Mechanics **188**, 133 (1988).

[41] N. Bentenitis and S. Krause, Droplet deformation in DC electric fields: The extended leaky dielectric model, Langmuir **21**, 6194 (2005).

[42] P. F. Salipante and P. M. Vlahovska, Electrohydrody-

namics of drops in strong uniform dc electric fields, Physics of Fluids **22**, 1 (2010).

[43] U. Ghosh and S. Chakraborty, Electrokinetics over charge-modulated surfaces in the presence of patterned wettability: Role of the anisotropic streaming potential, Physical Review E - Statistical, Nonlinear, and Soft Matter Physics **88**, 52 (2013).

[44] N.-T. Nguyen, Micro-magnetofluidics: Interactions between magnetism and fluid flow on the microscale, Microfluidics and Nanofluidics **12**, 1 (2012).

[45] R.-J. Yang, H.-H. Hou, Y.-N. Wang, and L.-M. Fu, Micro-magnetofluidics in microfluidic systems: A review, Sensors and Actuators, B: Chemical **224**, 1 (2016).

[46] N.-T. Nguyen, Deformation of ferrofluid marbles in the presence of a permanent magnet, Langmuir **29**, 13982 (2013).

[47] S. Torza, R. G. Cox, S. G. Mason, and G. I. Taylor, Electrohydrodynamic deformation and bursts of liquid drops, Philosophical Transactions of the Royal Society of London. Series A, Mathematical and Physical Sciences **269**, 295 (1971).

[48] P. A. Arp, R. T. Foister, and S. G. Mason, Some electrohydrodynamic effects in fluid dispersions, Advances in Colloid and Interface Science **12**, 295 (1980).

[49] C. Werner, R. Zimmermann, and T. Kratzmüller, Streaming potential and streaming current measurements at planar solid/liquid interfaces for simultaneous determination of zeta potential and surface conductivity, Colloids and Surfaces A: Physicochemical and Engineering Aspects **192**, 205 (2001).

[50] X. Chu and Y. Jian, Magnetohydrodynamic electroosmotic flow of Maxwell fluids with patterned charged surface in narrow confinements, Journal of Physics D: Applied Physics **52**, 10.1088/1361-6463/ab2b27 (2019).

[51] S. Santra, D. Sen, S. Das, and S. Chakraborty, Electrohydrodynamic interaction between droplet pairs in a confined shear flow, Physics of Fluids **31**, 10.1063/1.5088948 (2019).

[52] J.-W. Ha and S.-M. Yang, Rheological responses of oil-in-oil emulsions in an electric field, Journal of Rheology **44**, 235 (2000).

[53] W. J. Stanton, M. J. Etzel, and B. J. Walker, Electro-hydrodynamic Flow and Chaotic Mixing Inside Drops, UNIVERSITY OF CALIFORNIA Santa Barbara, , 634 (2007).

[54] X. Xu and G. M. Homsy, The settling velocity and shape distortion of drops in a uniform electric field, Journal of Fluid Mechanics **564**, 395 (2006).

[55] K. Stierstadt and M. Liu, Maxwell's stress tensor and the forces in magnetic liquids, ZAMM Zeitschrift fur Angewandte Mathematik und Mechanik **95**, 4 (2015).

[56] M. Liu and K. Stierstadt, *Electromagnetic Force and the Maxwell Stress Tensor in Condensed Systems*, January 1970 (2000) arXiv:0010261 [cond-mat].

[57] L. G. Leal, Advanced transport phenomena: Fluid mechanics and convective transport processes, Advanced Transport Phenomena: Fluid Mechanics and Convective Transport Processes **9780521849**, 1 (2007).

Pulak Gupta,[†] Purbarun Dhar,[‡] and Devranjan Samanta[†]

[*] Corresponding author: devranjan.samanta@iitrpr.ac.in

[†] Department of Mechanical Engineering, Indian Institute of Technology Ropar, Punjab–140001, India

[‡] Hydrodynamics and Thermal Multiphysics Lab (HTML), Department of Mechanical Engineering, Indian Institute of Technology Kharagpur, West Bengal–721302, India

| System B | $\sigma$(S/m) | $\varepsilon/\varepsilon_0$ | $\mu$(Kg/m$-$s) | $\rho$ (Kg/m$^3$) |
|---|---|---|---|---|
| Droplet (silicon oil) | $3.33 \times 10^{-11}$ | 2.77 | 12 | 980 |
| Suspending Medium (oxidized castor oil) | $1.11 \times 10^{-12}$ | 6.3 | 6.5 | 980 |

TABLE I. Physical properties of the fluids used. Here, , $R_{in}$ = 10 mm, $E_o$ = 10 V/mm, $\gamma = 5.5 \times 10^{-3}$ [$Nm^{-1}$] and $\epsilon_o = 8.854 \times 10^{-12}$ [$F$-$m^{-1}$]. System B correspond to silicon oil droplet surrounded in oxidized castor oil.

| System A (R<S, R<1, S>1) | | | | | | | | | | | | |
|---|---|---|---|---|---|---|---|---|---|---|---|---|
| | $\sigma_{rr}^{H} = \tau_{rr}^{H} - p$ | | | | | | | | | $\tau_{rr}^{E}$ | $(\sigma_{rr}^{H})_T + \tau_{rr}^{E}$ | | |
| $\alpha$ | B = -1.0T | | | B = 0.0T | | | B = +1.0T | | | | B = -1.0T | B = 0.0T | B = +1.0T |
| | $(\sigma_{rr}^{H})_E$ | $(\sigma_{rr}^{H})_M$ | $(\sigma_{rr}^{H})_T$ | $(\sigma_{rr}^{H})_E$ | $(\sigma_{rr}^{H})_M$ | $(\sigma_{rr}^{H})_T$ | $(\sigma_{rr}^{H})_E$ | $(\sigma_{rr}^{H})_M$ | $(\sigma_{rr}^{H})_T$ | | | | |
| **0.100** | -3.416 | -0.998 | -4.414 | -3.416 | 0.000 | -3.416 | -3.416 | 1.003 | -2.413 | -3.849 | -8.263 | -7.265 | -6.262 |
| **0.200** | -3.261 | -0.955 | -4.216 | -3.261 | 0.000 | -3.261 | -3.261 | 0.961 | -2.300 | -3.824 | -8.040 | -7.085 | -6.124 |
| **0.300** | -2.865 | -0.847 | -3.712 | -2.865 | 0.000 | -2.865 | -2.865 | 0.852 | -2.013 | -3.756 | -7.468 | -6.621 | -5.769 |
| **0.400** | -2.150 | -0.647 | -2.797 | -2.150 | 0.000 | -2.150 | -2.150 | 0.650 | -1.500 | -3.629 | -6.426 | -5.779 | -5.129 |
| **0.500** | -1.052 | -0.326 | -1.378 | -1.052 | 0.000 | -1.052 | -1.052 | 0.327 | -0.725 | -3.433 | -4.811 | -4.485 | -4.158 |
| **0.600** | 0.512 | 0.162 | 0.674 | 0.512 | 0.000 | 0.512 | 0.512 | -0.167 | 0.345 | -3.169 | -2.495 | -2.657 | -2.824 |
| **0.700** | 2.754 | 0.930 | 3.684 | 2.754 | 0.000 | 2.754 | 2.754 | -0.943 | 1.811 | -2.848 | 0.836 | -0.094 | -1.037 |
| **0.800** | 6.431 | 2.326 | 8.757 | 6.431 | 0.000 | 6.431 | 6.431 | -2.350 | 4.081 | -2.492 | 6.265 | 3.939 | 1.589 |
| **0.900** | 15.668 | 6.141 | 21.809 | 15.668 | 0.000 | 15.668 | 15.668 | -6.199 | 9.469 | -2.124 | 19.685 | 13.544 | 7.345 |

TABLE II. Steady State variation of Jump in normal electrical stress ($[\![\tau_{rr}^{E}]\!]$), jump in normal hydrodynamic stress due to electric field only ($[\![\sigma_{rr}^{H}]\!]_E$), jump in normal hydrodynamic stress due to EMHD coupling ($[\![\sigma_{rr}^{H}]\!]_M$), jump in total normal hydrodynamic stress (i.e. $[\![\sigma_{rr}^{H}]\!]_T = [\![\sigma_{rr}^{H}]\!]_E + [\![\sigma_{rr}^{H}]\!]_M$ ) and jump in total normal stress (i.e.$[\![\tau_{rr}^{T}]\!] = [\![\sigma_{rr}^{H}]\!]_E + [\![\sigma_{rr}^{H}]\!]_M + [\![\tau_{rr}^{E}]\!]$) with confinement ratio ($\alpha$) at B=-1.0, 0 and 1.0T for system A (R<S system).

| System B (R>S, R>1, S<1) | | | | | | | | | | | | |
|---|---|---|---|---|---|---|---|---|---|---|---|---|
| | $\sigma_{rr}^{H} = \tau_{rr}^{H} - p$ | | | | | | | | | $\tau_{rr}^{E}$ | $(\sigma_{rr}^{H})_T + \tau_{rr}^{E}$ | | |
| $\alpha$ | B = -1.0T | | | B = 0.0T | | | B = +1.0T | | | | B = -1.0T | B = 0.0T | B = +1.0T |
| | $(\sigma_{rr}^{H})_E$ | $(\sigma_{rr}^{H})_M$ | $(\sigma_{rr}^{H})_T$ | $(\sigma_{rr}^{H})_E$ | $(\sigma_{rr}^{H})_M$ | $(\sigma_{rr}^{H})_T$ | $(\sigma_{rr}^{H})_E$ | $(\sigma_{rr}^{H})_M$ | $(\sigma_{rr}^{H})_T$ | | | | |
| **0.100** | 0.206 | 2.156 | 2.362 | 0.206 | 0.000 | 0.206 | 0.206 | -2.156 | -1.950 | 3.967 | 6.329 | 4.173 | 2.017 |
| **0.200** | 0.205 | 2.126 | 2.331 | 0.205 | 0.000 | 0.205 | 0.205 | -2.127 | -1.922 | 4.018 | 6.349 | 4.223 | 2.096 |
| **0.300** | 0.201 | 2.047 | 2.248 | 0.201 | 0.000 | 0.201 | 0.201 | -2.048 | -1.847 | 4.161 | 6.409 | 4.362 | 2.314 |
| **0.400** | 0.191 | 1.883 | 2.074 | 0.191 | 0.000 | 0.191 | 0.191 | -1.883 | -1.692 | 4.463 | 6.537 | 4.654 | 2.771 |
| **0.500** | 0.167 | 1.549 | 1.716 | 0.167 | 0.000 | 0.167 | 0.167 | -1.549 | -1.382 | 5.038 | 6.754 | 5.205 | 3.656 |
| **0.600** | 0.098 | 0.826 | 0.924 | 0.098 | 0.000 | 0.098 | 0.098 | -0.825 | -0.727 | 6.125 | 7.049 | 6.223 | 5.398 |
| **0.700** | -0.137 | -0.986 | -1.123 | -0.137 | 0.000 | -0.137 | -0.137 | 0.986 | 0.849 | 8.345 | 7.222 | 8.208 | 9.194 |
| **0.800** | -1.215 | -6.814 | -8.029 | -1.215 | 0.000 | -1.215 | -1.215 | 6.814 | 5.599 | 13.805 | 5.776 | 12.590 | 19.404 |
| **0.900** | -11.051 | -39.245 | -50.296 | -11.051 | 0.000 | -11.051 | -11.051 | 39.244 | 28.193 | 34.507 | -15.789 | 23.456 | 62.700 |

TABLE III. Steady State variation of Jump in normal electrical stress ($[\![\tau_{rr}^{E}]\!]$), jump in normal hydrodynamic stress due to electric field only ($[\![\sigma_{rr}^{H}]\!]_E$), jump in normal hydrodynamic stress due to EMHD coupling ($[\![\sigma_{rr}^{H}]\!]_M$), jump in total normal hydrodynamic stress (i.e. $[\![\sigma_{rr}^{H}]\!]_T = [\![\sigma_{rr}^{H}]\!]_E + [\![\sigma_{rr}^{H}]\!]_M$ ) and jump in total normal stress (i.e.$[\![\tau_{rr}^{T}]\!] = [\![\sigma_{rr}^{H}]\!]_E + [\![\sigma_{rr}^{H}]\!]_M + [\![\tau_{rr}^{E}]\!]$) with confinement ratio ($\alpha$) at B=-1.0, 0 and 1.0T for system B (R>S system).

| t | $\tau_{rr}^{M}$ | $\tau_{rr}^{E}$ | $\sigma_{rr}^{H}$ | $p$ | $\tau_{rr}^{H}$ | $\tau_{rr}^{T}$ | $\tau_{rr}^{M}$ | $\tau_{rr}^{E}$ | $\sigma_{rr}^{H}$ | $p$ | $\tau_{rr}^{H}$ | $\tau_{rr}^{T}$ |
|---|---|---|---|---|---|---|---|---|---|---|---|---|
| | B = +1T, $\alpha$ = 0.2 | | | | | | B = +1T, $\alpha$ = 0.8 | | | | | |
| 0.01 | 1.36 | -0.14 | -1.20 | 1.23 | 0.03 | 0.01 | 0.98 | 0.57 | -1.54 | 1.55 | 0.01 | 0.00 |
| 0.12 | 0.29 | -3.07 | 2.67 | -3.13 | -0.45 | -0.11 | 0.10 | -2.27 | 2.17 | -2.47 | -0.30 | 0.00 |
| 0.23 | 0.06 | -3.67 | 3.20 | -3.86 | -0.67 | -0.41 | 0.01 | -2.47 | 2.47 | -2.83 | -0.36 | 0.01 |
| 0.34 | 0.01 | -3.79 | 3.04 | -3.78 | -0.74 | -0.74 | 0.00 | -2.49 | 2.51 | -2.87 | -0.36 | 0.02 |
| 0.45 | 0.00 | -3.82 | 2.76 | -3.54 | -0.77 | -1.05 | 0.00 | -2.49 | 2.52 | -2.88 | -0.36 | 0.03 |
| 0.56 | 0.00 | -3.82 | 2.47 | -3.27 | -0.80 | -1.35 | 0.00 | -2.49 | 2.53 | -2.89 | -0.36 | 0.04 |
| 0.67 | 0.00 | -3.82 | 2.19 | -3.02 | -0.82 | -1.63 | 0.00 | -2.49 | 2.54 | -2.90 | -0.36 | 0.05 |
| 0.78 | 0.00 | -3.82 | 1.93 | -2.78 | -0.85 | -1.89 | 0.00 | -2.49 | 2.54 | -2.90 | -0.36 | 0.06 |
| 0.89 | 0.00 | -3.82 | 1.68 | -2.55 | -0.87 | -2.14 | 0.00 | -2.49 | 2.55 | -2.91 | -0.36 | 0.06 |
| 1 | 0.00 | -3.82 | 1.45 | -2.33 | -0.89 | -2.38 | 0.00 | -2.49 | 2.56 | -2.92 | -0.36 | 0.07 |
| | **SYSTEM A** | | | | | | | | | | | |
| t | $\tau_{rr}^{M}$ | $\tau_{rr}^{E}$ | $\sigma_{rr}^{H}$ | $p$ | $\tau_{rr}^{H}$ | $\tau_{rr}^{T}$ | $\tau_{rr}^{M}$ | $\tau_{rr}^{E}$ | $\sigma_{rr}^{H}$ | $p$ | $\tau_{rr}^{H}$ | $\tau_{rr}^{T}$ |
| | B = 0T, $\alpha$ = 0.2 | | | | | | B = 0T, $\alpha$ = 0.8 | | | | | |
| 0.01 | 0.00 | -0.14 | 0.14 | -0.23 | -0.09 | 0.00 | 0.00 | 0.57 | -0.56 | 0.41 | -0.15 | 0.00 |
| 0.12 | 0.00 | -3.07 | 2.87 | -3.68 | -0.82 | -0.20 | 0.00 | -2.27 | 2.28 | -2.87 | -0.59 | 0.01 |
| 0.23 | 0.00 | -3.67 | 3.10 | -4.19 | -1.09 | -0.56 | 0.00 | -2.47 | 2.50 | -3.15 | -0.65 | 0.03 |
| 0.34 | 0.00 | -3.79 | 2.86 | -4.03 | -1.17 | -0.94 | 0.00 | -2.49 | 2.54 | -3.20 | -0.66 | 0.05 |
| 0.45 | 0.00 | -3.82 | 2.52 | -3.73 | -1.21 | -1.30 | 0.00 | -2.49 | 2.56 | -3.22 | -0.66 | 0.07 |
| 0.56 | 0.00 | -3.82 | 2.19 | -3.43 | -1.24 | -1.64 | 0.00 | -2.49 | 2.58 | -3.24 | -0.66 | 0.09 |
| 0.67 | 0.00 | -3.82 | 1.87 | -3.14 | -1.27 | -1.96 | 0.00 | -2.49 | 2.60 | -3.26 | -0.66 | 0.11 |
| 0.78 | 0.00 | -3.82 | 1.57 | -2.86 | -1.29 | -2.26 | 0.00 | -2.49 | 2.62 | -3.28 | -0.65 | 0.13 |
| 0.89 | 0.00 | -3.82 | 1.28 | -2.60 | -1.32 | -2.54 | 0.00 | -2.49 | 2.64 | -3.29 | -0.65 | 0.15 |
| 1 | 0.00 | -3.82 | 1.01 | -2.36 | -1.34 | -2.81 | 0.00 | -2.49 | 2.66 | -3.31 | -0.65 | 0.17 |
| | **SYSTEM A** | | | | | | | | | | | |
| t | $\tau_{rr}^{M}$ | $\tau_{rr}^{E}$ | $\sigma_{rr}^{H}$ | $p$ | $\tau_{rr}^{H}$ | $\tau_{rr}^{T}$ | $\tau_{rr}^{M}$ | $\tau_{rr}^{E}$ | $\sigma_{rr}^{H}$ | $p$ | $\tau_{rr}^{H}$ | $\tau_{rr}^{T}$ |
| | B = -1T, $\alpha$ = 0.2 | | | | | | B = -1T, $\alpha$ = 0.8 | | | | | |
| 0.01 | -1.36 | -0.14 | 1.49 | -1.70 | -0.21 | -0.01 | -0.98 | 0.57 | 0.41 | -0.73 | -0.32 | 0.00 |
| 0.12 | -0.29 | -3.07 | 3.06 | -4.24 | -1.18 | -0.30 | -0.10 | -2.27 | 2.39 | -3.26 | -0.87 | 0.02 |
| 0.23 | -0.06 | -3.67 | 3.01 | -4.52 | -1.50 | -0.71 | -0.01 | -2.47 | 2.53 | -3.48 | -0.95 | 0.05 |
| 0.34 | -0.01 | -3.79 | 2.67 | -4.27 | -1.60 | -1.13 | 0.00 | -2.49 | 2.57 | -3.53 | -0.96 | 0.08 |
| 0.45 | 0.00 | -3.82 | 2.28 | -3.93 | -1.65 | -1.54 | 0.00 | -2.49 | 2.60 | -3.56 | -0.95 | 0.12 |
| 0.56 | 0.00 | -3.82 | 1.90 | -3.59 | -1.69 | -1.92 | 0.00 | -2.49 | 2.64 | -3.59 | -0.95 | 0.15 |
| 0.67 | 0.00 | -3.82 | 1.54 | -3.26 | -1.72 | -2.28 | 0.00 | -2.49 | 2.67 | -3.62 | -0.95 | 0.18 |
| 0.78 | 0.00 | -3.82 | 1.20 | -2.95 | -1.74 | -2.62 | 0.00 | -2.49 | 2.70 | -3.65 | -0.95 | 0.21 |
| 0.89 | 0.00 | -3.82 | 0.88 | -2.65 | -1.77 | -2.94 | 0.00 | -2.49 | 2.73 | -3.68 | -0.95 | 0.24 |
| 1 | 0.00 | -3.82 | 0.58 | -2.38 | -1.79 | -3.24 | 0.00 | -2.49 | 2.76 | -3.71 | -0.94 | 0.27 |

TABLE IV. Transient variation of $\llbracket\tau_{rr}^{M}\rrbracket$, $\llbracket\tau_{rr}^{E}\rrbracket$, $\llbracket\sigma_{rr}^{H}\rrbracket(=\llbracket\tau_{rr}^{H}\rrbracket-\llbracket p\rrbracket)$, $\llbracket\tau_{rr}^{T}\rrbracket(=\llbracket\tau_{rr}^{M}\rrbracket+\llbracket\tau_{rr}^{E}\rrbracket+\llbracket\sigma_{rr}^{H}\rrbracket)$ at $\alpha$=0.2 and 0.8 at B=0, 1.0 and -1.0T for system A (R<S system)

| t | $\tau_{rr}^{M}$ | $\tau_{rr}^{E}$ | $\sigma_{rr}^{H}$ | $p$ | $\tau_{rr}^{H}$ | $\tau_{rr}^{T}$ | $\tau_{rr}^{M}$ | $\tau_{rr}^{E}$ | $\sigma_{rr}^{H}$ | $p$ | $\tau_{rr}^{H}$ | $\tau_{rr}^{T}$ |
|---|---|---|---|---|---|---|---|---|---|---|---|---|
| | B = +1T, $\alpha$ = 0.2 | | | | | | B = +1T, $\alpha$ = 0.8 | | | | | |
| 0.01 | -35.70 | 0.27 | 35.23 | -26.84 | 8.38 | -0.20 | -111.54 | 0.15 | 111.36 | -110.13 | 1.23 | -0.02 |
| 0.12 | -25.50 | 0.82 | 22.65 | -16.47 | 6.17 | -2.04 | -94.78 | 0.79 | 93.74 | -92.63 | 1.11 | -0.25 |
| 0.23 | -18.22 | 1.46 | 13.49 | -8.79 | 4.70 | -3.27 | -80.54 | 1.72 | 78.38 | -77.29 | 1.08 | -0.45 |
| 0.34 | -13.01 | 2.05 | 6.90 | -3.21 | 3.69 | -4.06 | -68.44 | 2.79 | 65.04 | -63.93 | 1.11 | -0.60 |
| 0.45 | -9.30 | 2.53 | 2.21 | 0.78 | 2.99 | -4.55 | -58.16 | 3.91 | 53.52 | -52.35 | 1.17 | -0.73 |
| 0.56 | -6.64 | 2.91 | -1.09 | 3.60 | 2.51 | -4.83 | -49.42 | 5.01 | 43.59 | -42.34 | 1.24 | -0.83 |
| 0.67 | -4.74 | 3.20 | -3.40 | 5.56 | 2.16 | -4.95 | -41.99 | 6.05 | 35.05 | -33.72 | 1.33 | -0.90 |
| 0.78 | -3.39 | 3.41 | -4.98 | 6.89 | 1.91 | -4.96 | -35.68 | 7.01 | 27.73 | -26.31 | 1.41 | -0.95 |
| 0.89 | -2.42 | 3.56 | -6.04 | 7.77 | 1.73 | -4.90 | -30.32 | 7.88 | 21.46 | -19.97 | 1.49 | -0.99 |
| 1 | -1.73 | 3.68 | -6.74 | 8.34 | 1.60 | -4.79 | -25.77 | 8.66 | 16.10 | -14.54 | 1.57 | -1.01 |
| | **SYSTEM B** | | | | | | | | | | | |
| t | $\tau_{rr}^{M}$ | $\tau_{rr}^{E}$ | $\sigma_{rr}^{H}$ | $p$ | $\tau_{rr}^{H}$ | $\tau_{rr}^{T}$ | $\tau_{rr}^{M}$ | $\tau_{rr}^{E}$ | $\sigma_{rr}^{H}$ | $p$ | $\tau_{rr}^{H}$ | $\tau_{rr}^{T}$ |
| | B = 0T, $\alpha$ = 0.2 | | | | | | B = 0T, $\alpha$ = 0.8 | | | | | |
| 0.01 | 0.00 | 0.27 | -0.27 | 0.26 | -0.01 | 0.00 | 0.00 | 0.15 | -0.15 | 0.14 | -0.01 | 0.00 |
| 0.12 | 0.00 | 0.82 | -0.79 | 0.61 | -0.18 | 0.03 | 0.00 | 0.79 | -0.79 | 0.62 | -0.17 | 0.00 |
| 0.23 | 0.00 | 1.47 | -1.37 | 1.17 | -0.20 | 0.10 | 0.00 | 1.72 | -1.72 | 1.49 | -0.23 | 0.00 |
| 0.34 | 0.00 | 2.06 | -1.88 | 1.72 | -0.16 | 0.18 | 0.00 | 2.79 | -2.79 | 2.57 | -0.23 | 0.00 |
| 0.45 | 0.00 | 2.55 | -2.27 | 2.17 | -0.10 | 0.28 | 0.00 | 3.90 | -3.90 | 3.72 | -0.19 | 0.00 |
| 0.56 | 0.00 | 2.93 | -2.55 | 2.50 | -0.05 | 0.38 | 0.00 | 5.01 | -4.99 | 4.86 | -0.13 | 0.01 |
| 0.67 | 0.00 | 3.22 | -2.73 | 2.72 | -0.01 | 0.50 | 0.00 | 6.05 | -6.02 | 5.96 | -0.06 | 0.02 |
| 0.78 | 0.00 | 3.44 | -2.83 | 2.85 | 0.02 | 0.61 | 0.00 | 7.01 | -6.97 | 6.97 | 0.01 | 0.04 |
| 0.89 | 0.00 | 3.60 | -2.87 | 2.92 | 0.04 | 0.73 | 0.00 | 7.88 | -7.82 | 7.90 | 0.08 | 0.06 |
| 1 | 0.00 | 3.71 | -2.88 | 2.93 | 0.06 | 0.84 | 0.00 | 8.66 | -8.58 | 8.72 | 0.15 | 0.08 |
| | **SYSTEM B** | | | | | | | | | | | |
| t | $\tau_{rr}^{M}$ | $\tau_{rr}^{E}$ | $\sigma_{rr}^{H}$ | $p$ | $\tau_{rr}^{H}$ | $\tau_{rr}^{T}$ | $\tau_{rr}^{M}$ | $\tau_{rr}^{E}$ | $\sigma_{rr}^{H}$ | $p$ | $\tau_{rr}^{H}$ | $\tau_{rr}^{T}$ |
| | B = -1T, $\alpha$ = 0.2 | | | | | | B = -1T, $\alpha$ = 0.8 | | | | | |
| 0.01 | 35.70 | 0.27 | -35.76 | 27.35 | -8.41 | 0.21 | 111.54 | 0.15 | -111.67 | 110.41 | -1.26 | 0.02 |
| 0.12 | 25.50 | 0.82 | -24.22 | 17.68 | -6.54 | 2.10 | 94.78 | 0.79 | -95.31 | 93.87 | -1.45 | 0.25 |
| 0.23 | 18.22 | 1.46 | -16.23 | 11.14 | -5.09 | 3.45 | 80.54 | 1.72 | -81.81 | 80.28 | -1.54 | 0.44 |
| 0.34 | 13.01 | 2.05 | -10.66 | 6.66 | -4.01 | 4.40 | 68.44 | 2.79 | -70.63 | 69.07 | -1.56 | 0.60 |
| 0.45 | 9.30 | 2.53 | -6.76 | 3.55 | -3.20 | 5.07 | 58.16 | 3.91 | -61.33 | 59.78 | -1.55 | 0.74 |
| 0.56 | 6.64 | 2.91 | -4.00 | 1.39 | -2.61 | 5.55 | 49.42 | 5.01 | -53.57 | 52.07 | -1.51 | 0.85 |
| 0.67 | 4.74 | 3.20 | -2.06 | -0.13 | -2.18 | 5.88 | 41.99 | 6.05 | -47.09 | 45.64 | -1.45 | 0.95 |
| 0.78 | 3.39 | 3.41 | -0.68 | -1.19 | -1.87 | 6.12 | 35.68 | 7.01 | -41.66 | 40.26 | -1.39 | 1.03 |
| 0.89 | 2.42 | 3.56 | 0.29 | -1.94 | -1.65 | 6.28 | 30.32 | 7.88 | -37.09 | 35.76 | -1.33 | 1.11 |
| 1 | 1.73 | 3.68 | 0.98 | -2.47 | -1.49 | 6.39 | 25.77 | 8.66 | -33.26 | 31.98 | -1.28 | 1.17 |

TABLE V. Transient variation of $\llbracket\tau_{rr}^{M}\rrbracket$, $\llbracket\tau_{rr}^{E}\rrbracket$, $\llbracket\sigma_{rr}^{H}\rrbracket(=\llbracket\tau_{rr}^{H}\rrbracket-\llbracket p\rrbracket)$, $\llbracket\tau_{rr}^{T}\rrbracket(=\llbracket\tau_{rr}^{M}\rrbracket+\llbracket\tau_{rr}^{E}\rrbracket+\llbracket\sigma_{rr}^{H}\rrbracket)$ at $\alpha$=0.2 and 0.8 at B=0, 1.0 and -1.0T for system B (R>S system)

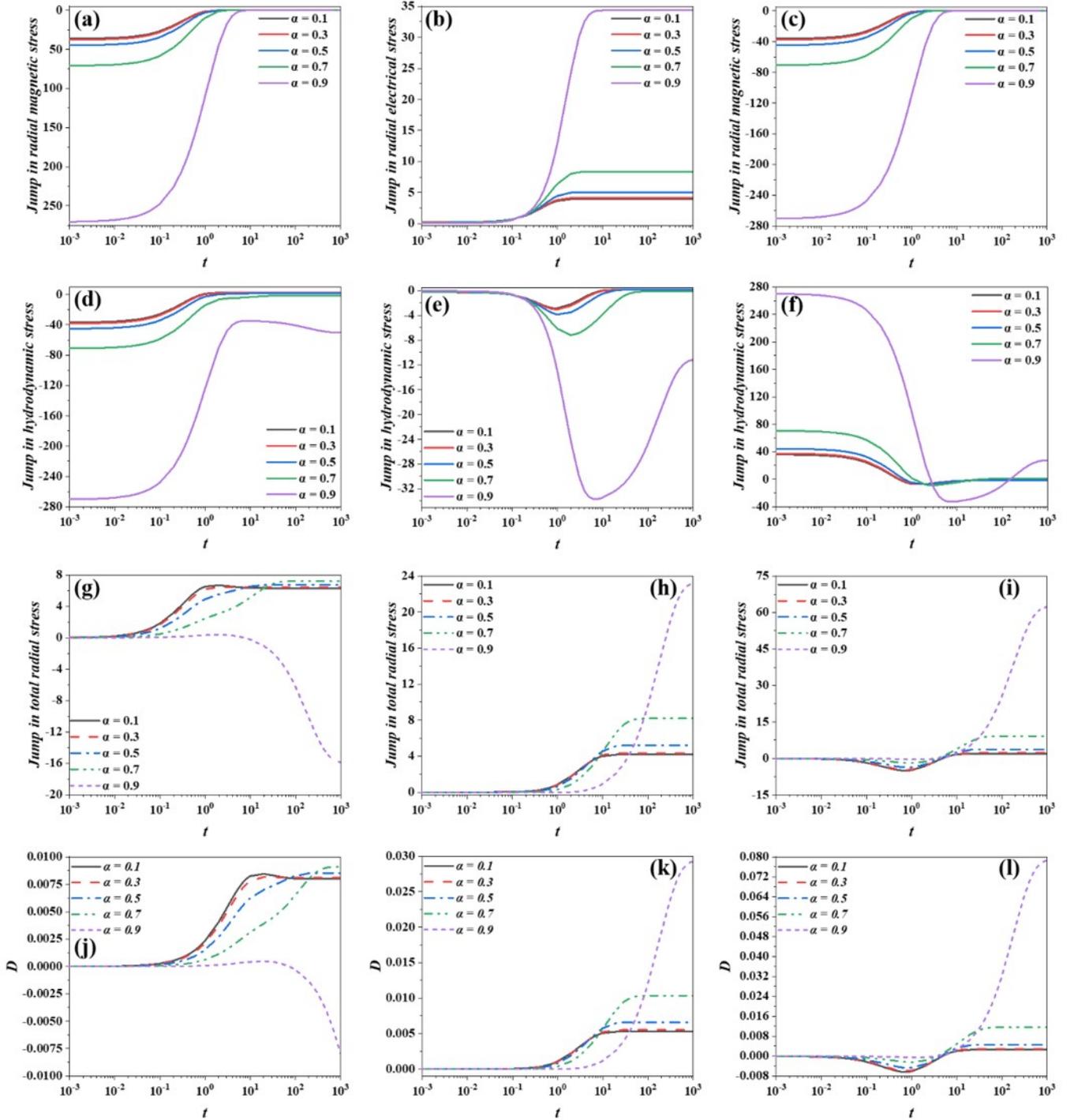

FIG. 1. Transient variation of $[\![\tau^M_{rr}]\!]$ at different $\alpha$ at constant $E_o = 10$ V/mm for system B at (a) B = -1.0 T and (c) B = +1.0T; (b) Variation of $[\![\tau^E_{rr}]\!]$ for B=0T; Variation of $[\![\sigma^H_{rr}]\!](=[\![\tau^H_{rr}]\!]-[\![\,p\,]\!])$ at (d) B = -1.0T, (e) B = 0.0T, (f) B = +1.0T; variation of $[\![\tau^T_{rr}]\!]$ at (g) B = -1.0T, (h) B = 0.0T, (i) B = +1.0T. The time history evolution of $D$ at different $\alpha$ at constant $E_o = 10$ V/mm for system B at (j) B = -1.0T, (k) B = 0.0T, (l) B = +1.0T.

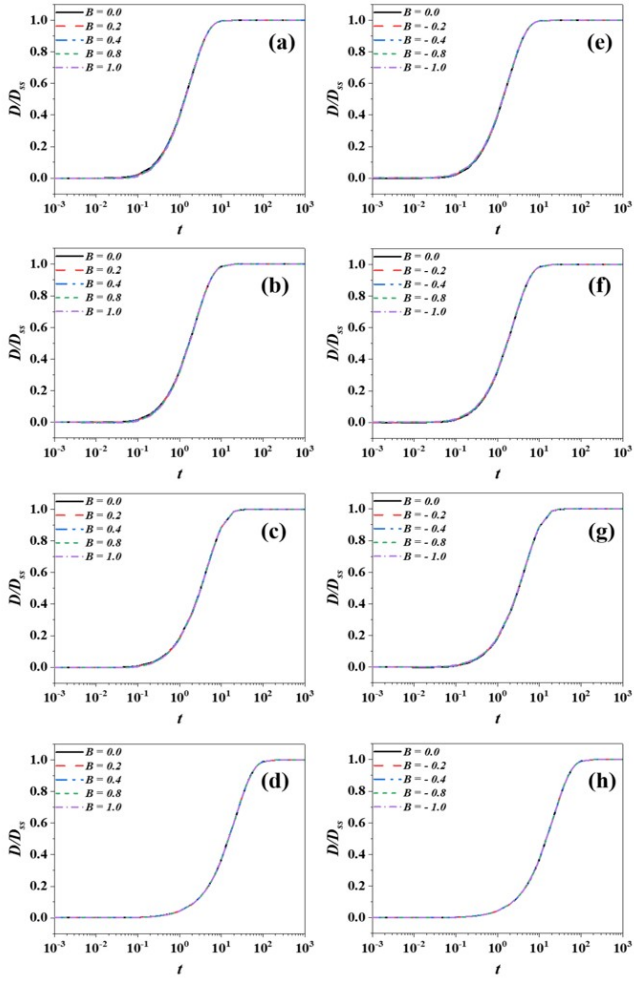

FIG. 2. Transient evolution of $D/D_{ss}$ for different B for system A at constant $E_o = 10$ V/mm i.e. (a) +B, $\alpha = 0.2$, (b) +B, $\alpha = 0.4$, (c) +B, $\alpha = 0.6$, (d) +B, $\alpha = 0.8$, (e) -B, $\alpha = 0.2$, (f) -B, $\alpha = 0.4$, (g) -B, $\alpha = 0.6$, (h) -B, $\alpha = 0.8$.

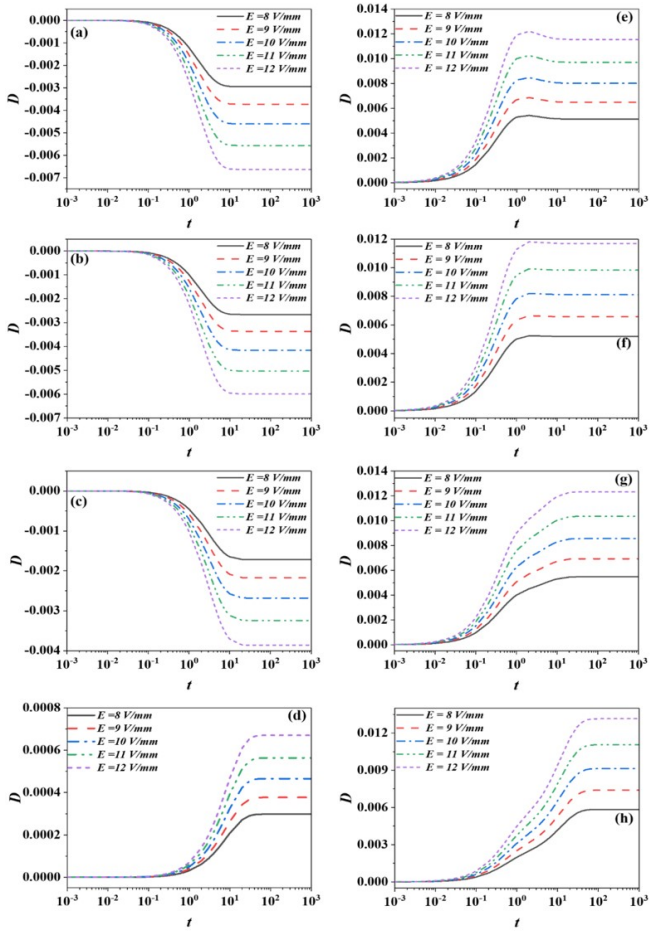

FIG. 3. Temporal deformation of $D$ for different electric field at a constant magnetic field strength (B = -1.0 T) for system A with (a) $\alpha = 0.1$, (b) $\alpha = 0.3$, (c) $\alpha = 0.5$, (d) $\alpha = 0.7$ and for system B with (e) $\alpha = 0.1$, (f) $\alpha = 0.3$, (g) $\alpha = 0.5$, (h) $\alpha = 0.7$.

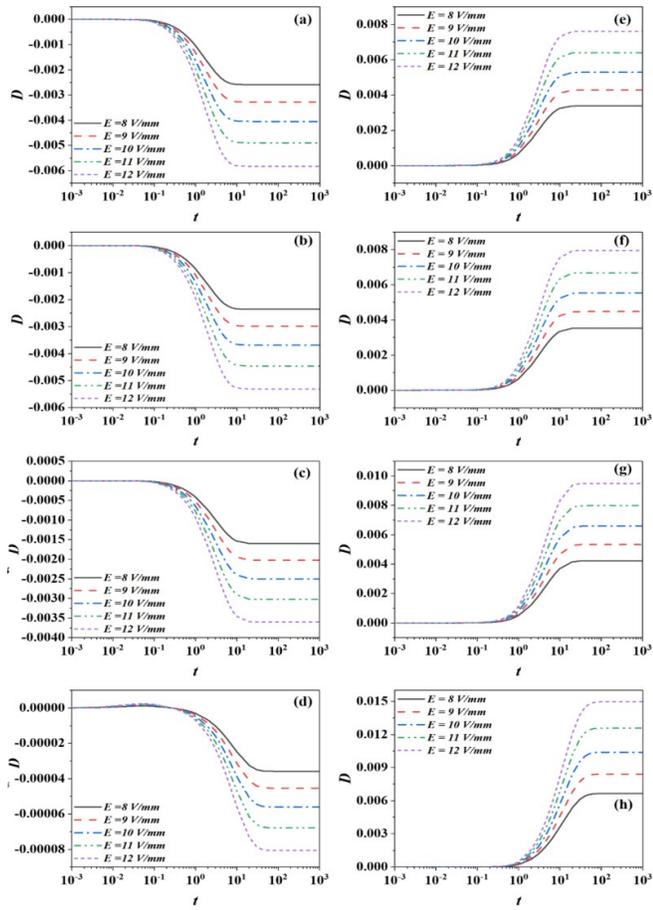

FIG. 4. Temporal deformation of $D$ for different electric field and at B = 0.0 T for system A with ratio (a) $\alpha = 0.1$, (b) $\alpha = 0.3$, (c) $\alpha = 0.5$, (d) $\alpha = 0.7$ and for system B with (e) $\alpha = 0.1$, (f) $\alpha = 0.3$, (g) $\alpha = 0.5$, (h) $\alpha = 0.7$.

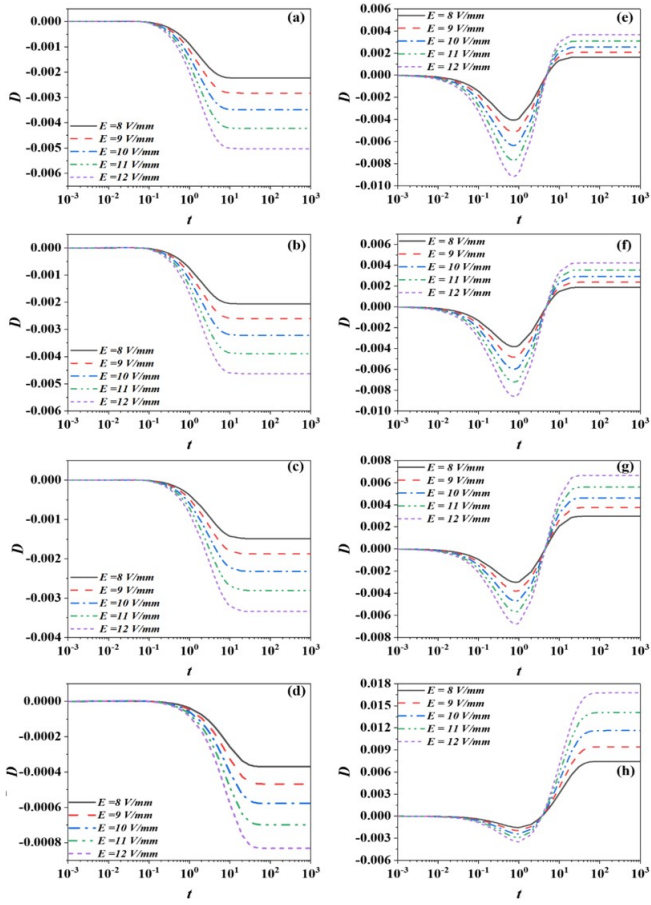

FIG. 5. Temporal deformation of $D$ for different electric field at a constant magnetic field strength ( B = +1.0 T) for system A with (a) $\alpha = 0.1$, (b) $\alpha = 0.3$, (c) $\alpha = 0.5$, (d) $\alpha = 0.7$ and for system B with (e) $\alpha = 0.1$, (f) $\alpha = 0.3$, (g) $\alpha = 0.5$, (h) $\alpha = 0.7$.

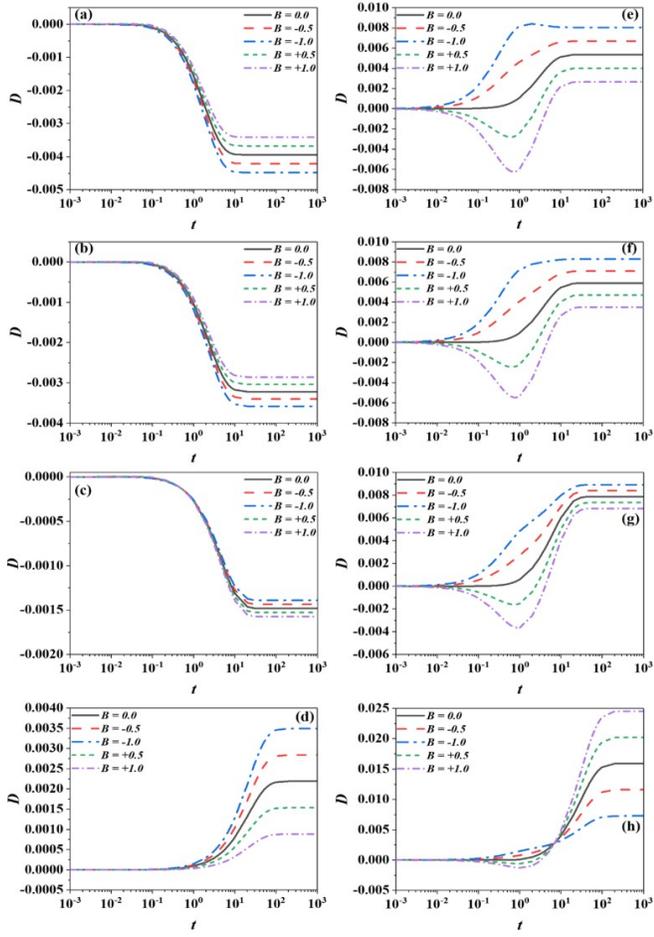

FIG. 6. Temporal deformation of $D$ for different B at constant $E_o = 10$ V/mm for system A with (a) $\alpha = 0.2$, (b) $\alpha = 0.4$, (c) $\alpha = 0.6$, (d) $\alpha = 0.8$ and for system B with (e) $\alpha = 0.2$, (f) $\alpha = 0.4$, (g) $\alpha = 0.6$, (h) $\alpha = 0.8$.

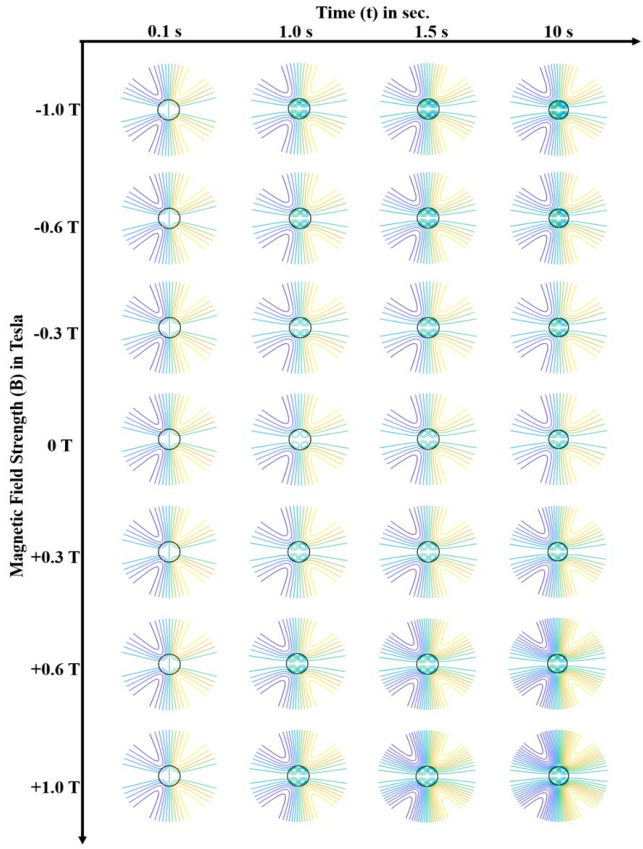

FIG. 7. Transient evolution of streamline pattern for different B at constant $E_o = 10$ V/mm at $\alpha = 0.0$ for system A from table 1. i.e. oxidized castor oil in silicon oil. The non-dimensional numbers are Oh = 51.69, Ca = 0.00446, R = 0.033, S = 2.27, $\lambda = 0.542$. Here outside fluid is more conductive than droplet liquid (i.e. R < S, R < 1, $\lambda$ < 1).

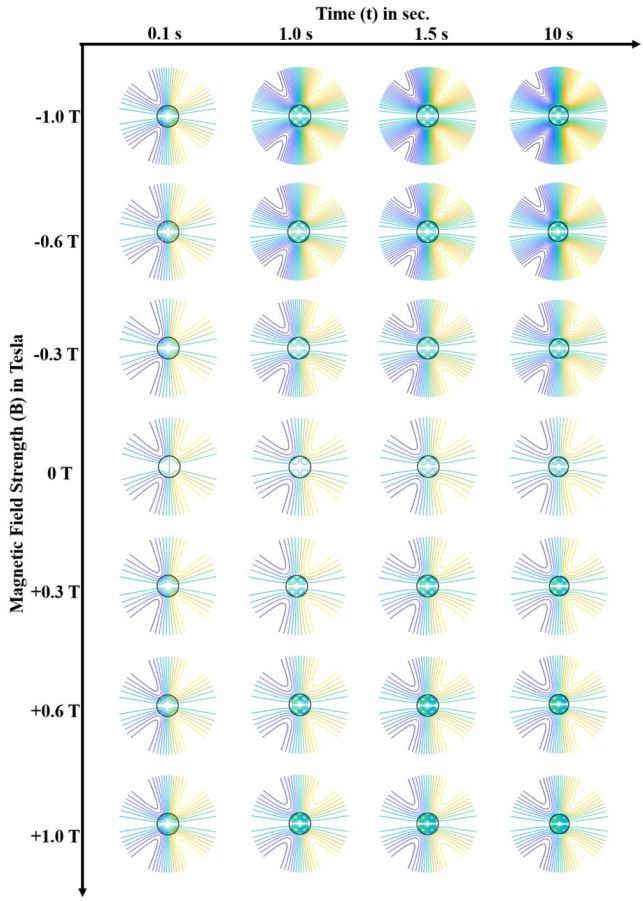

FIG. 8. Transient evolution of streamline pattern for different B at constant $E_o = 10$ V/mm at $\alpha = 0.0$ for system B from table 1. i.e. silicon oil in oxidized castor oil. The non-dimensional numbers are Oh = 28, Ca = 0.0101, R = 30, S = 0.44, $\lambda = 1.846$. Here droplet liquid is more conductive than outside fluid (i.e. R > S, R > 1, $\lambda > 1$).

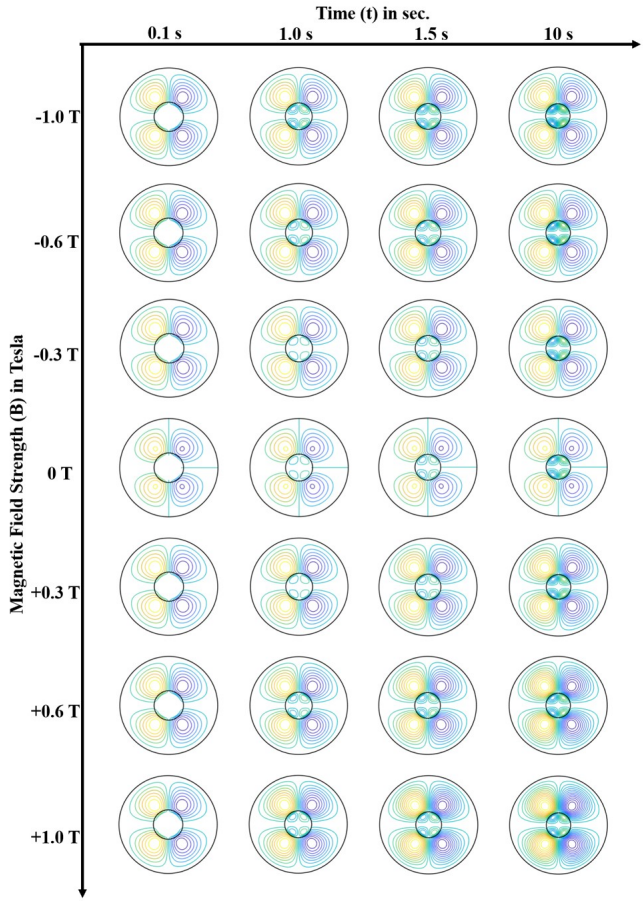

FIG. 9. Transient evolution of streamline pattern for different B at constant $E_o = 10$ V/mm at $\alpha = 0.2$ for system A i.e. oxidized castor oil in silicon oil (R < S, R < 1, $\lambda$ < 1).

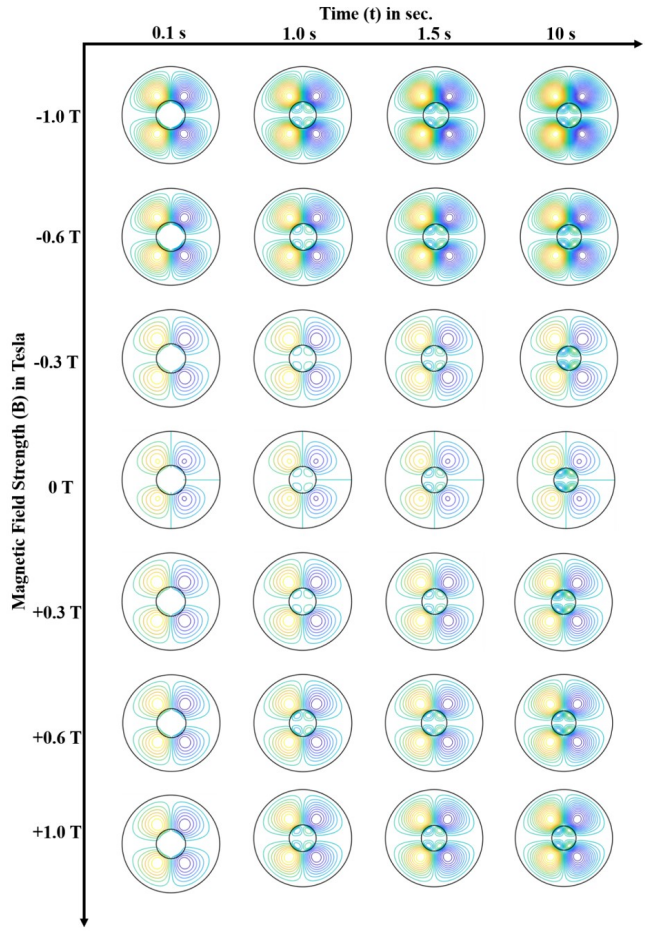

FIG. 10. Transient evolution of streamline pattern for different B at constant $E_o = 10$ V/mm at $\alpha = 0.2$ for system B i.e. silicon oil in oxidized castor oil (R > S, R > 1, $\lambda$ > 1).

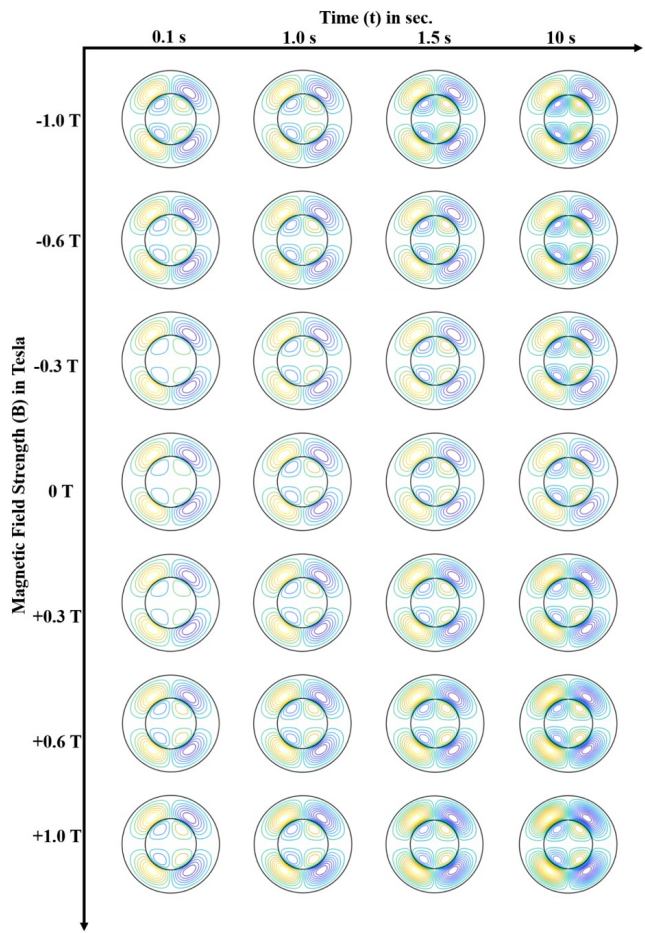

FIG. 11. Transient evolution of streamline pattern for different B at constant $E_o = 10$ V/mm at $\alpha = 0.4$ for system A i.e. oxidized castor oil in silicon oil (R < S, R < 1, $\lambda$ < 1).

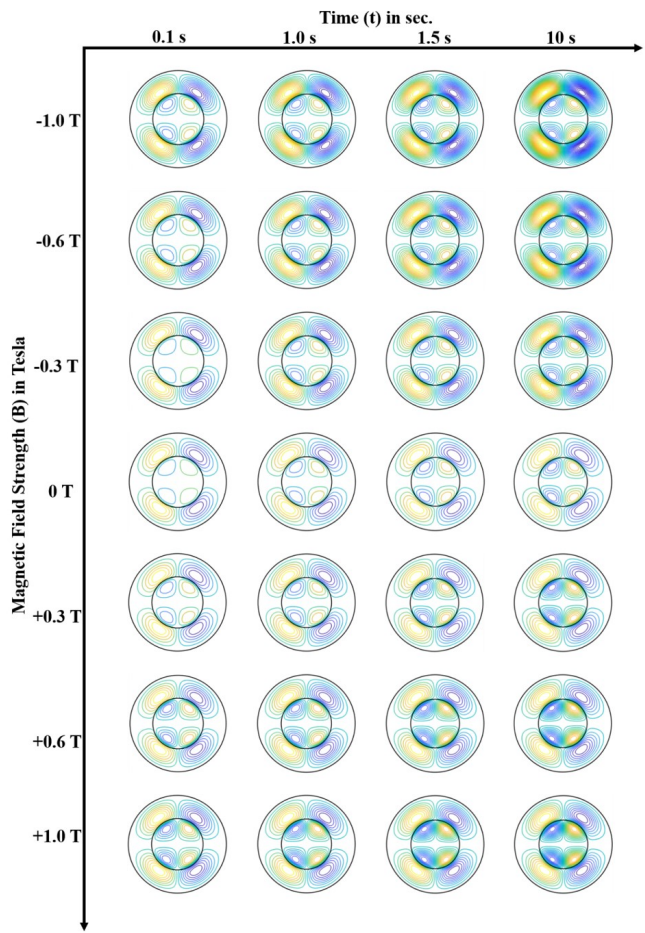

FIG. 12. Transient evolution of streamline pattern for different B at constant $E_o = 10$ V/mm at $\alpha = 0.4$ for system B i.e. silicon oil in oxidized castor oil (R > S, R > 1, $\lambda$ > 1).

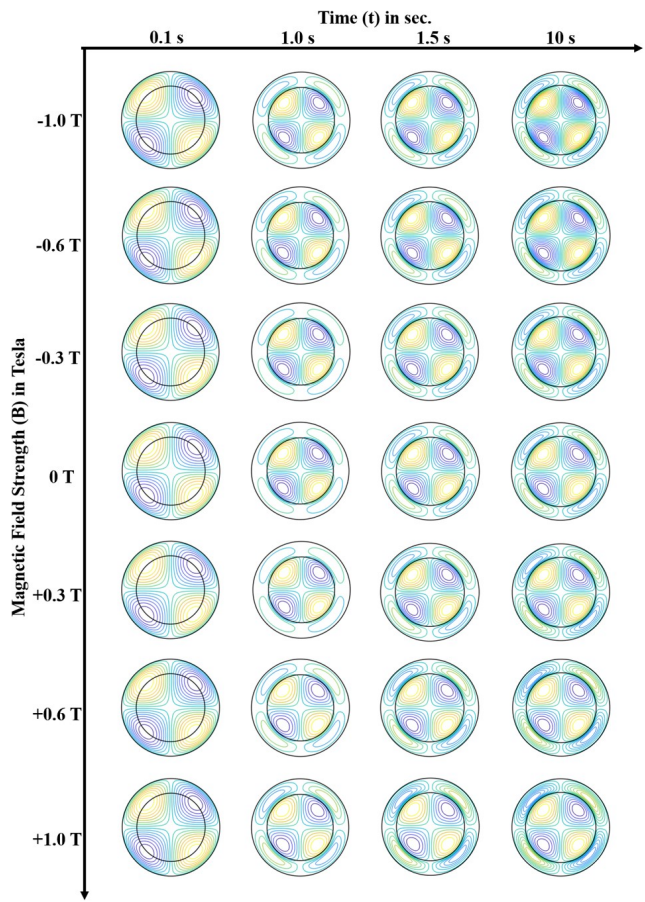

FIG. 13. Transient evolution of streamline pattern for different B at constant $E_o = 10$ V/mm at $\alpha = 0.7$ for system A i.e. oxidized castor oil in silicon oil (R < S, R < 1, $\lambda$ < 1).

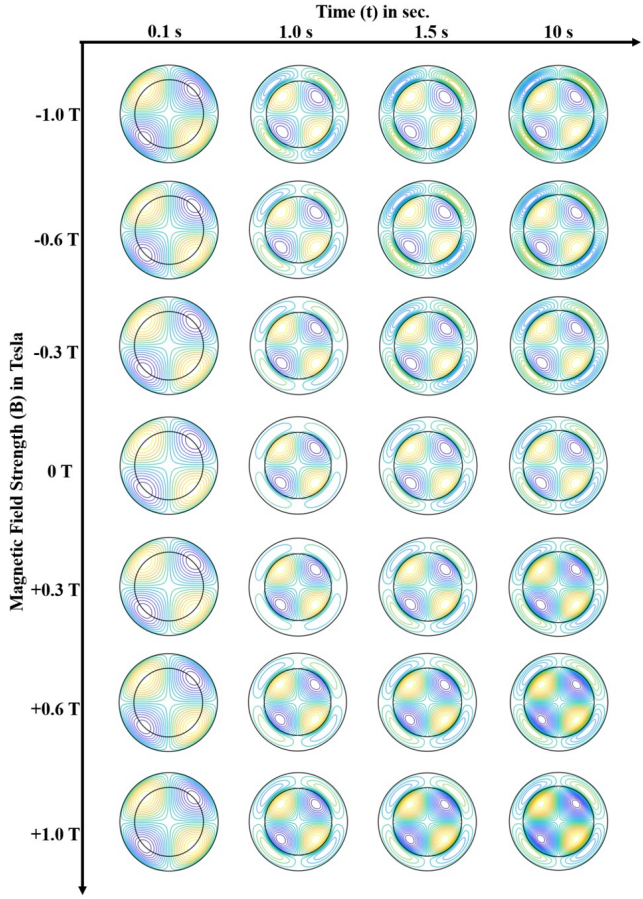

FIG. 14. Transient evolution of streamline pattern for different B at constant $E_o = 10$ V/mm at $\alpha = 0.7$ for system B i.e. silicon oil in oxidized castor oil (R > S, R > 1, $\lambda$ > 1).

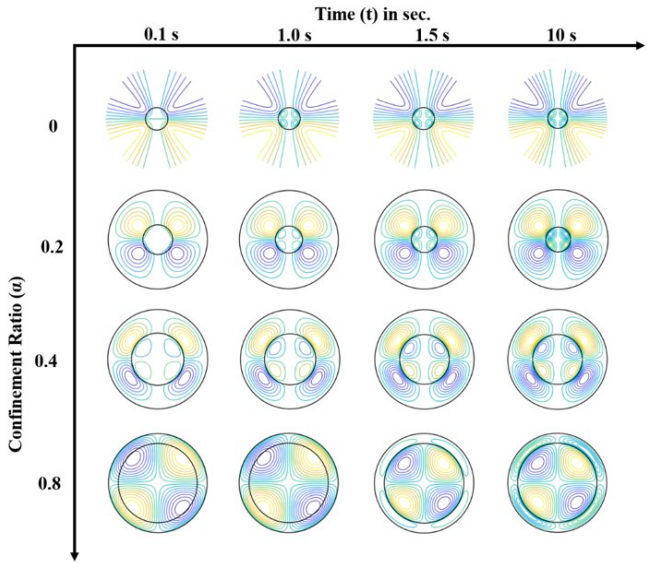

FIG. 15. Transient evolution of streamline pattern for different $\alpha$ at constant $E_o = 10$ V/mm at B = +0.3 T for system A i.e. oxidized castor oil in silicon oil (R < S, R < 1, $\lambda$ < 1).

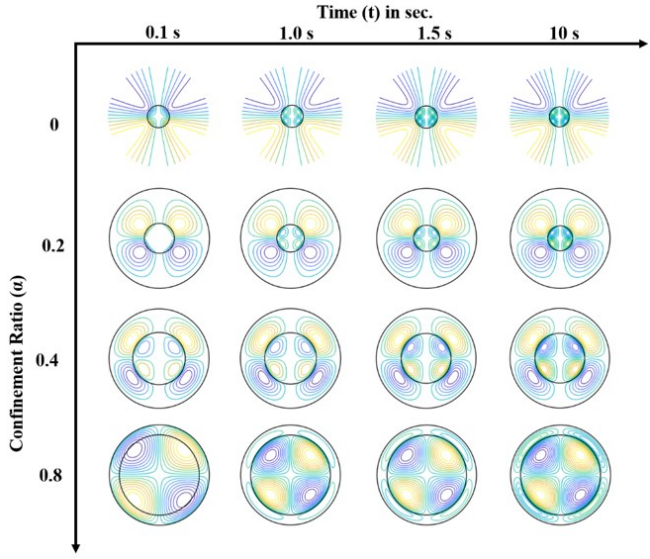

FIG. 16. Transient evolution of streamline pattern for different $\alpha$ at constant $E_o = 10$ V/mm at B = +0.3 T for system B i.e. silicon oil in oxidized castor oil (R > S, R > 1, $\lambda$ > 1).

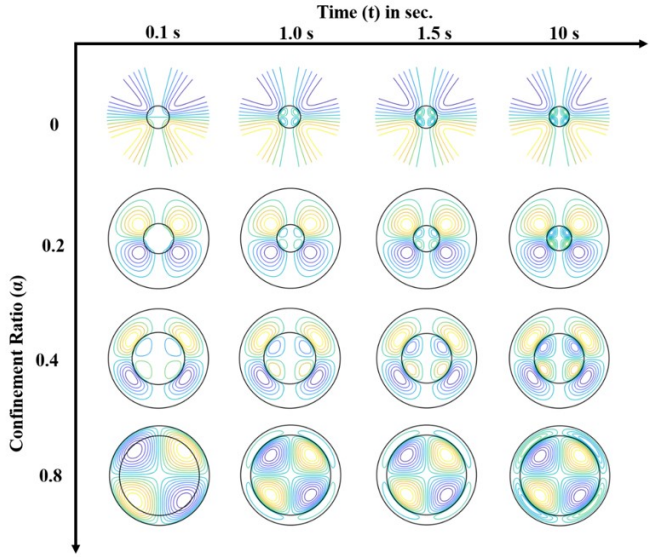

FIG. 17. Transient evolution of streamline pattern for different $\alpha$ at constant $E_o = 10$ V/mm at B = -0.3 T for system A i.e. oxidized castor oil in silicon oil (R < S, R < 1, $\lambda$ < 1).

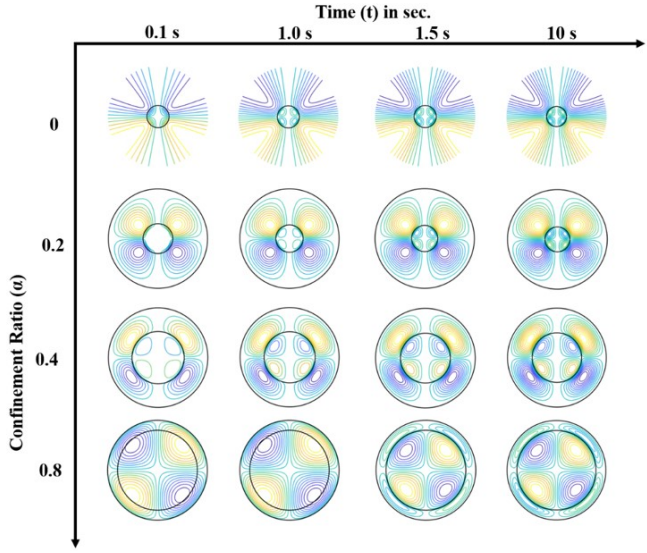

FIG. 18. Transient evolution of streamline pattern for different $\alpha$ at constant $E_o = 10$ V/mm at B = -0.3 T for system B i.e. silicon oil in oxidized castor oil (R > S, R > 1, $\lambda$ > 1).

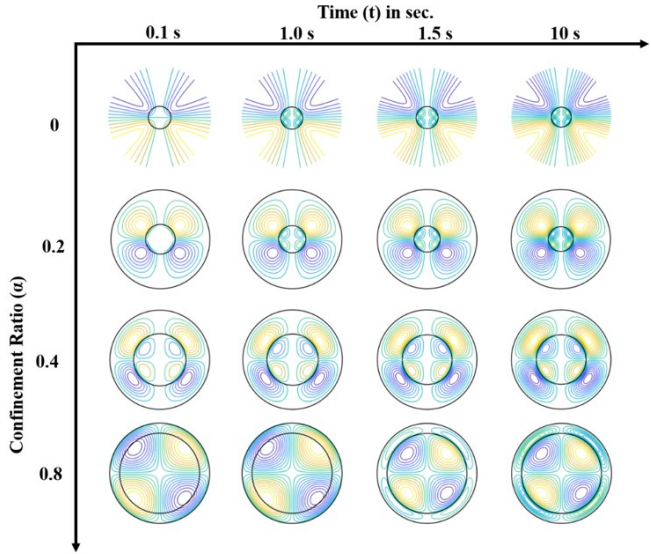

FIG. 19. Transient evolution of streamline pattern for different $\alpha$ at constant $E_o = 10$ V/mm at B = +0.6 T for system A i.e. oxidized castor oil in silicon oil (R < S, R < 1, $\lambda$ < 1).

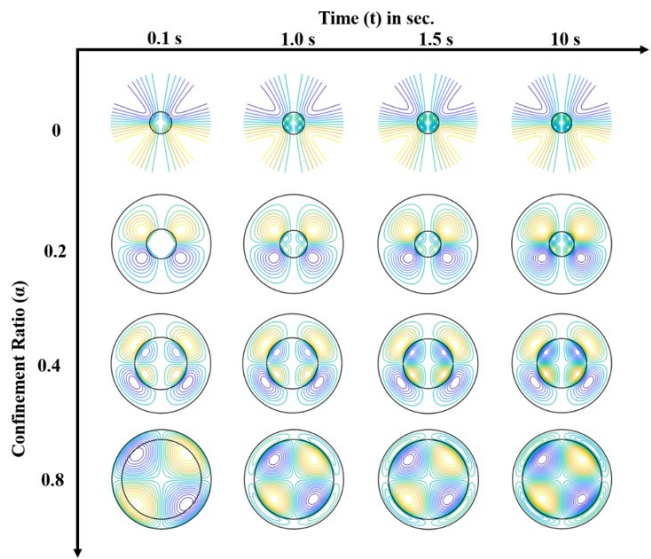

FIG. 20. Transient evolution of streamline pattern for different $\alpha$ at constant $E_o = 10$ V/mm at B = +0.6 T for system B i.e. silicon oil in oxidized castor oil (R > S, R > 1, $\lambda > 1$).

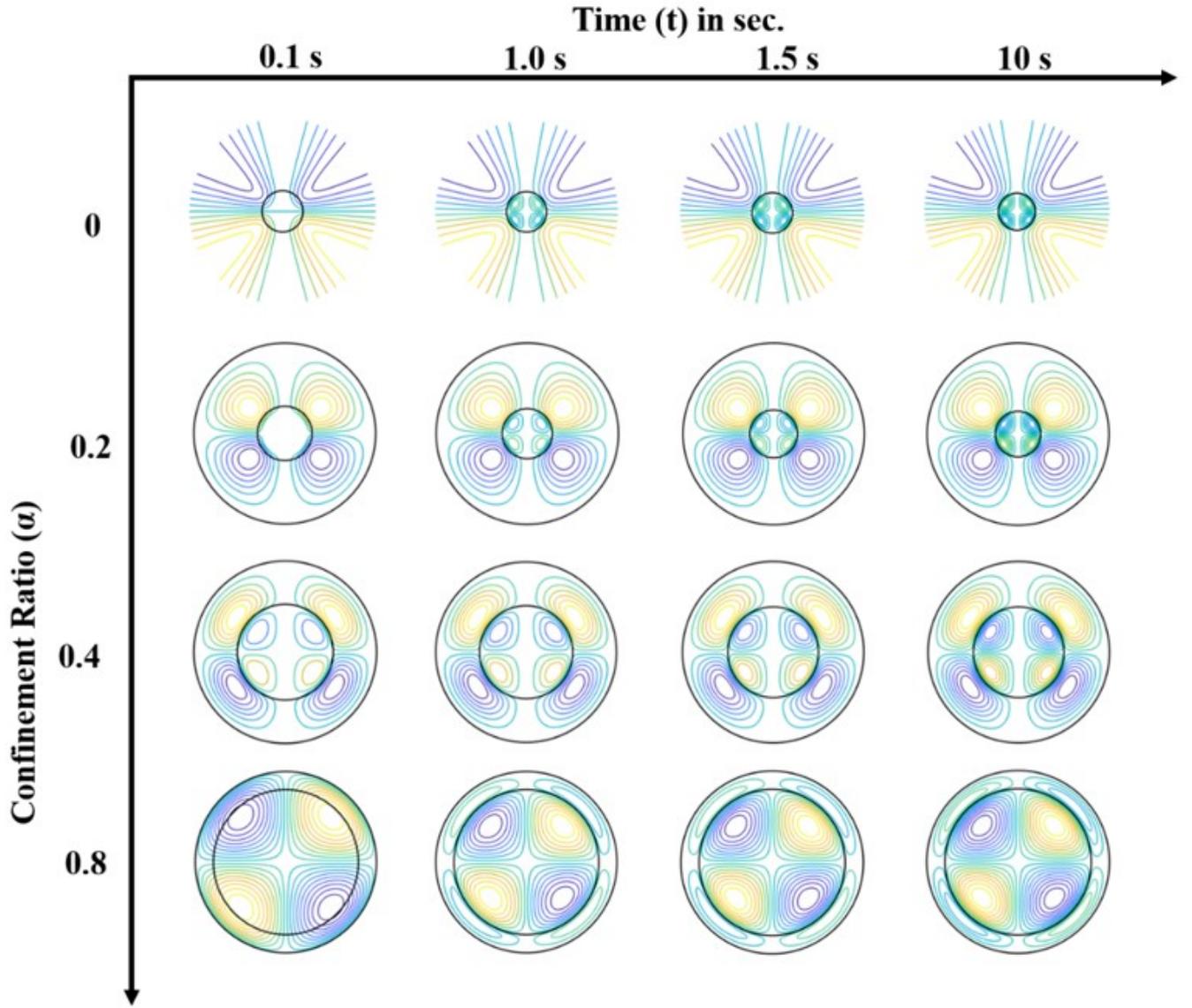

FIG. 21. Transient evolution of streamline pattern for different $\alpha$ at constant $E_o = 10$ V/mm at B = -0.6 T for system A i.e. oxidized castor oil in silicon oil (R < S, R < 1, $\lambda$ < 1).

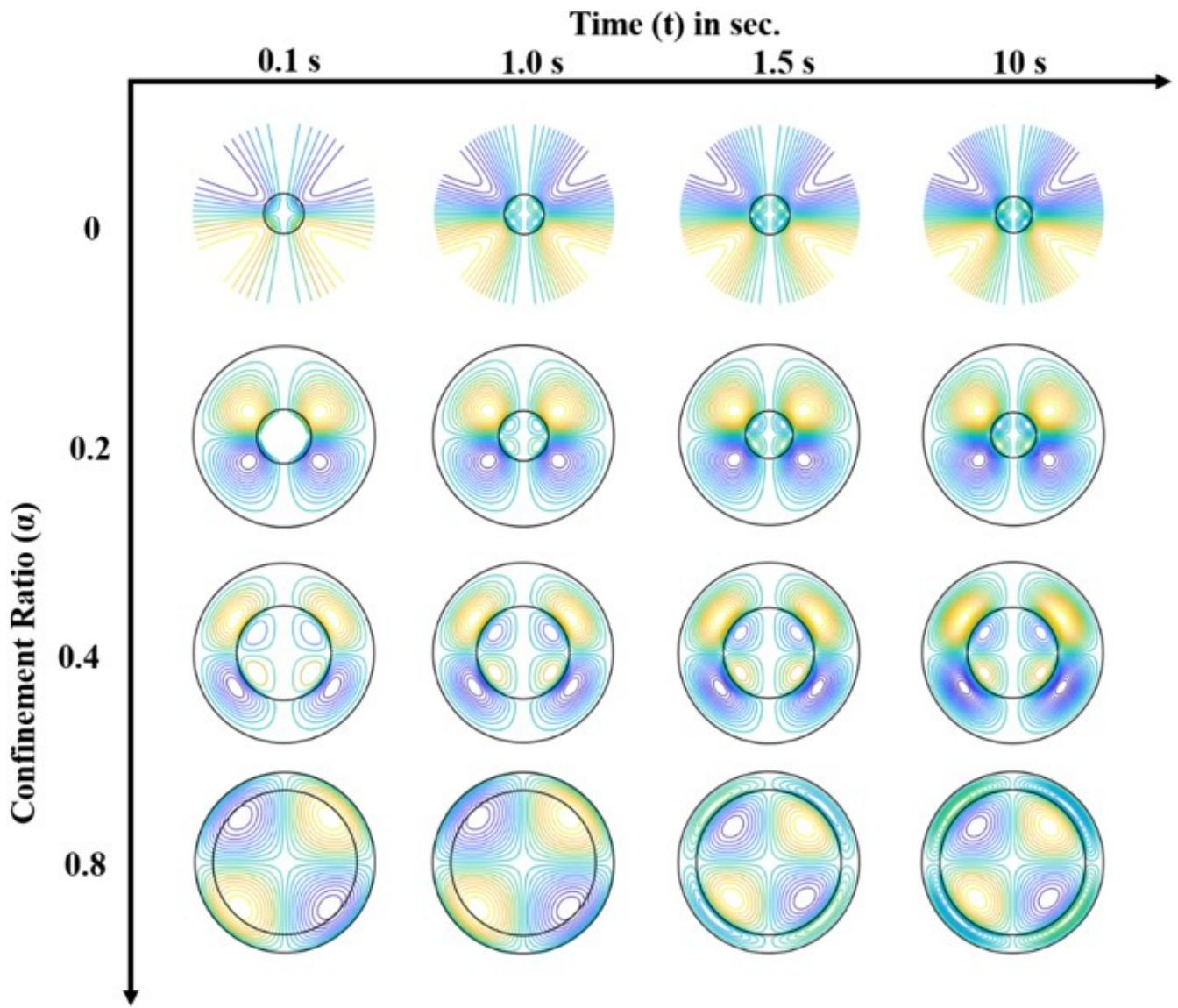

FIG. 22. Transient evolution of streamline pattern for different $\alpha$ at constant $E_o = 10$ V/mm at B = -0.6 T for system B silicon oil in oxidized castor oil (R > S, R > 1, $\lambda$ > 1)